# Nanotube-based one-dimensional heterostructures coupled by van der Waals forces

*Sofie Cambré,\* Ming Liu, Dmitry Levshov, Keigo Otsuka, Shigeo Maruyama,\* and Rong Xiang\**

Prof. S. Cambré, Dr. D. Levshov
Nanostructured and Organic Optical and Electronic Materials,
Department of Physics,
University of Antwerp, Belgium
E-mail: sofie.cambre@uantwerpen.be

Dr. M. Liu, Prof. K. Otsuka, Prof. S. Maruyama, Prof. R. Xiang
Department of Mechanical Engineering,
The University of Tokyo,
Tokyo 113-8656, Japan
E-mail: maruyama@photon.t.u-tokyo.ac.jp; xiangrong@photon.t.u-tokyo.ac.jp

##All authors contributed equally to this work

**Abstract**

One-dimensional (1D) van der Waals heterostructures based on carbon nanotube templates are raising a lot of excitement due to the possibility of creating new optical and electronic properties, by either confining molecules inside their hollow core or by adding layers on the outside of the nanotube. In contrast to their 2D analogues, where the number of layers, atomic type and relative orientation of the constituting layers are the main parameters defining physical properties, 1D heterostructures provide an additional degree of freedom, *i.e*. their specific diameter and chiral structure, for engineering their characteristics. This review discusses the current state-of-the-art in synthesizing 1D heterostructures, in particular focusing on their resulting optical properties, and details the vast parameter space that can be used to design heterostructures with custom-built properties that can be integrated into a large variety of applications. The review starts from describing the effects of van der Waals coupling on the properties of the simplest and best-studied 1D heterostructure, namely a double-wall carbon nanotube, then considers heterostructures built from the inside and the outside, which all use a nanotube as a template, and, finally, provides an outlook for the future of this research field.

## 1. Introduction

By reducing at least one characteristic dimension of a material to the nanometer scale, quantum effects and an increased surface-to-volume ratio lead to a dramatic change of the optical, electronic, mechanical and thermal properties. A particular interest in this perspective is the combination of two or more nanomaterials where the functionalities of the individual building blocks couple and yield new, fundamentally different properties, enabling a wide range of applications in diverse fields as chemistry, physics, biology, medicine and engineering. Particularly exciting in this context are carbon nanostructures, that allow tuning the dimensions of such heterostructures into two-dimensions (2D, *e.g.*, stacking of multiple layered materials on top of each other[1,2]), one-dimension (1D, using a nanotube as a template[3,4]) or even zero-dimensions (based on fullerenes[5] and carbon nano-onions[6,7]. In this review, we focus on carbon nanotubes (CNTs), mostly single-walled carbon nanotubes (SWCNTs), as a template for 1D heterostructures. In theory, such a 1D heterostructure can be formed in different manners either by employing the hollow inner space of the SWCNTs to encapsulate various molecules or by externally wrapping the SWCNT template with additional atomic layers. Figure 1 presents a selected overview of the 1D heterostructures discussed in this review.

A SWCNT can be conceptualized as a cylindrical structure that originates from rolling up a one-atom-thick single layer of graphene into a hollow cylinder.[8] SWCNTs combine a high mechanical resilience and chemical resistance, with unique and remarkably diverse optical and electronic properties (*e.g.*, metallic or semiconducting behavior) that depend critically on their exact diameter and chiral structure, defined by the chiral indices ($n,m$).[9,10] The 1D geometry and the associated quantum confinement leads to van Hove singularities in their electronic density of states, with characteristic optical transitions between these singularities in the visible through mid-IR.[9] The reduced dielectric screening in 1D furthermore leads to the formation of strongly bound excitons even observable at room temperature (binding energies of the order of several hundred meV).[11] Besides, SWCNTs possess smooth and impermeable walls, and their diameter can be tuned with subatomic precision by selecting a specific chiral structure. These combined properties make SWCNTs one of the most versatile and controlled 1D templates to date for synthesizing new 1D heterostructures. Moreover, as by definition, all carbon atoms of a SWCNT reside at its surface, the electronic, optical and vibrational properties of these 1D materials are extremely sensitive to their external and internal environment. Thus, the design of new heterostructures based on SWCNTs is often guided by improving their already peculiar optical properties.

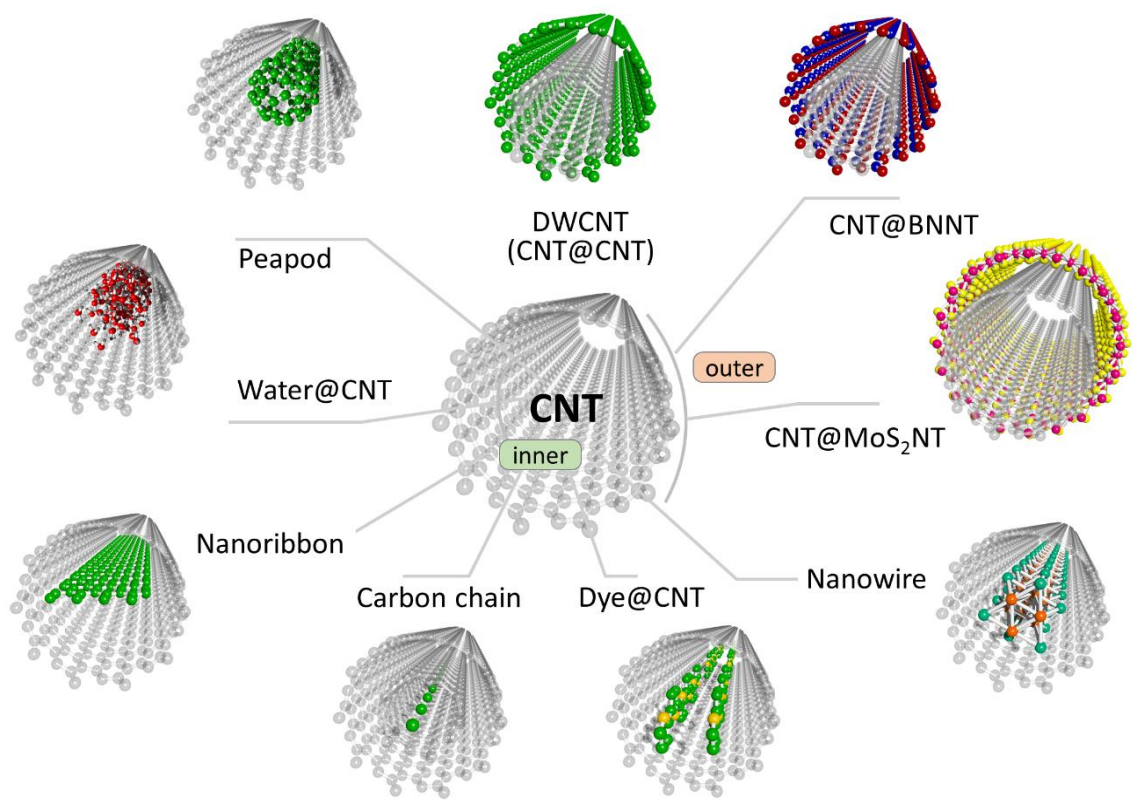

**Figure 1** A variety of nanotube-based 1D heterostructures fabricated so far using either the inner space or the outer surface of the CNT template. Representative structures are DWCNTs (CNT@CNT, where the "@" indicates encapsulation), peapods (fullerene@CNT), water@CNT, graphene-nanoribbon@CNT, linear-carbon-chain@DWCNT, dye@CNT, nanowire@CNT, CNT@BNNT and CNT@MoS$_2$NT.

The model textbook example of such a 1D SWCNT-based heterostructure is a double-wall carbon nanotube (DWCNT). It is composed of two coaxially stacked SWCNTs and inherits many of their fascinating physical properties. Interestingly, although DWCNTs are homogeneously carbon, they are often viewed as a distinct novel material instead of a simple sum of two SWCNTs since they possess remarkable features induced exclusively by van der Waals (vdW) interlayer coupling. The first observation of a DWCNT was reported in the seminal paper of Iijima in 1991[12], followed by a long series of extensive theoretical and experimental research still active today. Early spectroscopic studies focused mainly on macroscopic samples of DWCNTs in the form of bundles or non-sorted dispersions with a non-negligible percentage of SWCNT and multi-walled CNT (MWCNT) impurities. They showed that DWCNTs conserve most SWCNT-specific features, *e.g.*, the antenna (depolarization) effect and the presence of van Hove singularities,[13] but also provide unprecedented shielding of the inner layers' phonon properties from the environment.[14] Uncovering the vdW-induced effects required the separation of particular DWCNT configurations and was pursued in two independent ways. In 2009, Hersam *et al.* attained the sorting of macroscopic DWCNT samples from SWCNT impurities by density gradient ultracentrifugation (DGU) and later sorting by the outer-layer electronic type[15,16]; more recently, Flavel *et al.* achieved the separation of all DWCNT electronic configurations.[17] An alternative research route focusing on microscopic individual CNTs was first introduced by the group of M. Dresselhaus. It consisted in probing individual DWCNTs deposited on substrates and evidenced the inner pressure effects in DWCNTs.[18] A significant step forward consisted in preparing dedicated samples of horizontally aligned DWCNTs suspended over (high-resolution transmission electron microscopy) HRTEM grids or long hollow slits, which enabled simultaneous optical studies and structural characterization by TEM of the same suspended DWCNTs free of strong environmental interactions. It led to the most significant discoveries of the effect of vdW coupling on DWCNT properties.[19–25]

This vdW coupling can affect the heterostructure properties both from the "inside" and "outside". For instance, throughout the years, the hollow inner space of SWCNTs has generated much creativity in designing new heterostructures by encapsulation of various materials. In this case, the SWCNT provides an exceptionally stable environment for the encapsulated systems, and the number of molecules that have been encapsulated is endless, starting with two observations in 1998. Firstly, it was observed that after selectively opening the ends of the SWCNTs using HCl, Ru metal could enter the SWCNTs.[26] In the same year, $C_{60}$ fullerenes were found inside the hollow core of SWCNTs by TEM[27] and were considered by-products of the synthesis of SWCNTs by laser vaporization. Shortly after, metallofullerenes were encapsulated for the first time[28], and it was demonstrated that they could modulate the bandgap of the SWCNTs through the exerted strain, creating so-called peapod-like structures that could potentially be used to divide the elongated SWCNT structure into short segments with quantum-dot like behavior.[29] However, even though these early studies are numerous, it took until 2003 that this research field really started to bloom, namely when T. Takenobu *et al.* [30] demonstrated that by filling SWCNTs with electron donor or acceptor molecules, a stable and amphoteric doping of semiconducting SWCNTs could be achieved. This demonstration emphasized the potential of tuning the SWCNTs' properties by encapsulating various molecules inside their hollow core. The simplicity of the filling procedure, *i.e.,* plainly exposing pre-opened SWCNTs to the vapor phase of the molecules, combined with the versatility of the doping (choosing a molecule with a different electron affinity or ionization energy) and the unprecedented stability of the doping (protection by the SWCNT walls) opened up a new route towards the synthesis of macroscopic samples of filled SWCNTs, of great importance for future applications.

However, compared to the large number of 1D heterostructures formed by encapsulation, the heterostructures built on the outer surface of a SWCNT are much less represented in literature. The early activities were stimulated by the need for dispersing SWCNTs in a solvent, where usually a monolayer of surfactant or DNA molecules were adsorbed densely on the surface of an isolated SWCNT.[31–34] The presence of these molecules on the outer SWCNT surface can slightly affect the physical properties of the SWCNTs, but usually only served to keep the SWCNTs isolated in suspension. For decades, the only observed crystalized tubular material on a CNT was hexagonal boron nitride (h-BN),[35] that has the same honeycomb-like atomic arrangement as CNTs and a very similar lattice constant. The first synthesis of CNT@BNNT (CNT wrapped by BNNT) coaxial heterostructures was accidental; polyhedral and tubular graphitic nanoparticles made of separated carbon layers and BN layers were observed using TEM by Suenaga *et al.*,[36] in the soot collected on the anode deposit formed by arc discharging. Later, researchers developed a more intentional and designed synthesis route and confirmed that BNNT coating on multi-wall CNTs (MWCNTs) could preserve or even improve the properties of the original MWCNTs.[37] More recently, with the rise of 2D materials, a group of atomic-thin crystals started to emerge. Various function-designed transition metal dichalcogenide (TMDC) based heterostructures were synthesized around MWCNTs, and used for sensing or electro-chemical applications.[38,39] However, in these cases, TMDC nanotubes seem to form only around MWCNTs. This is likely associated with the TMDC's three-atomic-layer geometry and thus higher stiffness, which makes the formation of small diameter tubular structures less energetically-favorable.[40] Last year, a highly crystallized single-walled $MoS_2$ nanotube ($MoS_2$NT) was successfully synthesized around a SWCNT. The resulting CNT@$MoS_2$NT double-walled heteronanotubes have a diameter of only 4-5 nm and can be viewed as radial semiconductor-semiconductor junctions, which is an ideal structure to study the vdW coupling in tubular heterostructures.[3] CNT@BNNT@$MoS_2$NT 1D vdW heterostructures were also demonstrated. The intermediate BNNT can further serve as an additional parameter to tune the degree of coupling between the inner-most SWCNT and the outer-most $MoS_2$NT.[41]

In this review, we will first use DWCNTs as a textbook model to introduce the current understanding of the inter-tube vdW coupling in the simplest 1D heterostructure. Then, we present various more sophisticated SWCNT-based tubular heterostructures, formed by either encapsulating foreign molecules/atoms into the inner channel of SWCNTs or by constructing layers on the highly curved outer SWCNT surface. Particular emphasis will be given to the synthesis methodology and the optical properties of these tubular 1D heterostructures. We will demonstrate how this weak vdW coupling can affect physical properties of the stacking components or even result in new materials that cannot be synthesized without a SWCNT template. Lastly, an outlook is presented to picture the developing trend of this research field and highlight several important and interesting future research directions.

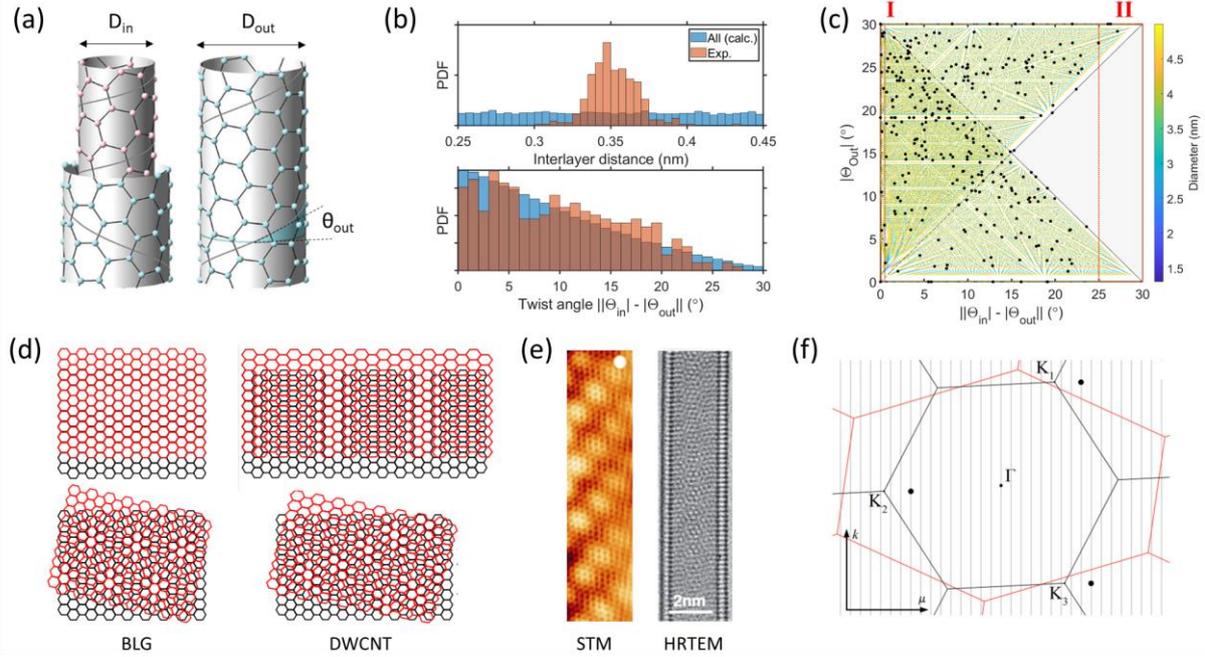

Figure 2. **Geometry of a DWCNT and experimentally observed DWCNT configurations**. (a) Geometrical model of a DWCNT, indicating diameter, chiral angle and handedness of the layers. Adapted with permission from [[42]]. Copyright 2019, Royal Society of Chemistry. (b)-(c) Statistical analysis of 335 DWCNTs, index-identified by ED or HRTEM, from about 20 papers (Adapted with permission from [[42]]; Copyright 2019, Royal Society of Chemistry): (b) Distributions of interlayer distance $\Delta r$ (top) and twist angle $\Delta\theta$ (bottom) for experimentally observed DWCNTs (red bars) compared with a random distribution of twist angles for all possible $(n_{in}, m_{in})@(n_{out}, m_{out})$ combinations in the ranges $1.0 < D_{out} < 5.0$ nm and $0.30 < \Delta r < 0.40$ nm (blue bars). (c) Plot of the outer layer's chiral angle versus the interlayer twist angle for the experimental (black dots) and calculated (colored dots) data from (b). The color bar indicates the outer diameters (in nm) of the DWCNTs in the calculated random distribution. Regions I and II refer to DWCNT twist angles $\Delta\theta \approx 0°$ and $\Delta\theta > 25°$, respectively, considered forbidden in ref. [[25]] due to interlayer coupling, but later found in other studies.[42,43] (d) Comparison of moiré patterns in AA-stacked bilayer graphene and DWCNTs, emphasizing the non-equivalence of the two systems due to the curvature effect (adapted with permission from [[44]]. Copyright 2016, Nature Publishing Group.). (e) Visualization of moiré patterns in DWCNTs by STM (adapted with permission from [[45]]. Copyright 2013, Elsevier) and HRTEM (adapted with permission from [[25]]. Copyright 2017, American Chemical Society). (f) The reciprocal space of (16,9)@(24,10) DWCNT, showing the effective compression and stretching of 2D graphene Brillouin zones due to curvature (adapted with permission from ref. [[46]] Copyright (2020) by the American Physical Society).

## 2. Van-der-Waals Interlayer coupling in DWCNTs

The long history of experimental and theoretical research of DWCNTs makes them an ideal model system for introducing vdW coupling in 1D. DWCNTs provide valuable insights into the effects of vdW interaction on the heterostructure growth, stability and physical properties. For instance, the vdW interaction between layers of non-commensurate DWCNTs was for a long time considered weak compared to the effects of strong covalent bonding within the layer of an SWCNT. However, recent experimental observations supported by theoretical studies have revealed, in such coaxial systems, substantial modulation of electronic and phonon properties due to intertube coupling[20,22,23,46,47]. The latter was found to depend on the geometry of the DWCNTs (Figure 2a), *i.e.*, the relative position of the layers and the way carbon atoms are arranged at their surfaces. This geometry is uniquely characterized by the chiral indices $(n_{in},m_{in})@(n_{out},m_{out})$ of the constituent inner and outer SWCNTs; or alternatively by DWCNT system parameters such as the average diameter $<D> = (D_{out} + D_{in})/2$, interlayer distance $\Delta r = (D_{out} - D_{in})/2$, and the interlayer twist angle $\Delta\theta = ||\theta_{in}|-|\theta_{out}||$.

The analysis of the statistical distributions of structural parameters for naturally-grown DWCNTs highlights the role of vdW coupling in the stability and energetics of the heterostructures. As an illustration, we review here 335 DWCNTs from the literature, synthesized by chemical vapor deposition (CVD), arc-discharge, and peapod conversion techniques (red bars and black dots in Figure 2b and Figure 2c, respectively) and unambiguously characterized by electron diffraction (ED) or HRTEM (statistics are based on the overview provided in Ref.[42], which holds most references but with several additional DWCNTs added from Refs.[22,43,48]). We compare the experimental statistics with the calculated set of all geometrically possible DWCNTs. The latter corresponds to any inner-outer SWCNT pairs falling within an outer diameter range of $1.0 < D_{out} < 5.0$ nm and an interlayer distance range of $0.30 < \Delta r < 0.40$ nm (blue bars and colored dots in Figures 2b,c, respectively), where the ranges of $D_{out}$ and $\Delta r$ correspond to the limits of the experimental distributions[42].

The interlayer distance $\Delta r$ in naturally-grown DWCNTs is distributed mainly between 0.30 – 0.40 nm with a peak value at around 0.34-0.35 nm (Figure 2b, top panel). It contrasts with the constant $\Delta r$ distribution for the set of all geometrically-possible DWCNT combinations (blue bars in Figure 2b). These experimental observations can be explained by theoretical studies of DWCNT energetics. For instance, Saito *et al.* showed that the DWCNT total energy primarily depends on the interlayer distance and has a minimum at around 0.33 – 0.35 nm due to vdW interaction[49]. Later, Guo *et al*. investigated the dependence on the twist angle $\Delta\theta$ and found several energetically preferable combinations. However, the energy difference between them and other combinations is relatively small and can play a role only during low-temperature growth[50]. Indeed, the experimentally observed twist angle distribution in Figure 2b (bottom panel) shows minor deviations from the calculated one, where all chiral combinations are taken (red versus blue bars in Figure 2b) since most of the analyzed DWCNTs were grown at high temperatures. Another problem was raised in 2017 by Ghedjatti *et al.,* who hypothesized that the formation of specific DWCNTs with $\Delta\theta \approx 0°$ and $\Delta\theta > 25°$ (regions I and II in Figure 2c, respectively) is forbidden due to the mechanical vdW coupling. [25] However, later independent studies by He *et al.*[43] and Rochal *et al.*[42] demonstrated that these geometrical configurations do exist.

One of the most critical effects of the vdW coupling concerns the electronic properties of DWCNTs, which is related to the concept of the moiré pattern, *i.e.* the superstructure obtained from the superposition of two periodic lattices. There is a fundamental difference between moiré patterns in 1D and 2D materials, as shown in Figure 2d[44]. For instance, in bilayer graphene (BLG), a moiré pattern can be only observed when $\Delta\theta \neq 0$ (Figure 2d, bottom left) and is absent in the so-called AA- and AB-stacked BLG ($\Delta\theta = 0$: Figure 2d, top left). In contrast, moiré patterns exist both in commensurate and non-commensurate DWCNTs due to the curvature effects and were observed experimentally by low-temperature (scanning tunnel microscopy) STM[45] and HRTEM[25] (Figure 2e). Note that the moiré pattern in a HRTEM image results from the superposition of four twisted graphene layers (at the top and bottom of a DWCNT), and not two, as in Figure 2d.

Several theoretical attempts studying incommensurate DWCNTs suggested that inter-tube electronic coupling is negligible between incommensurate inner- and outer-wall carbon lattices because couplings at different carbon atom sites oscillate with random phases and cancel each other out[51]. Although this is mostly true for random pairs of electronic states, there are important exceptions: an outer-tube state $\beta(q')$ can have finite electronic coupling to an inner-tube state $\alpha(q)$, *e.g.*, representing the $E_{ii}$ transition in the inner tube, if it satisfies the requirement $q' = q + G_i$, where $G_i$ is the reciprocal lattice vector of the inner tube's 2D graphene Brillouin zone. Such coupled states are designated by solid dots in the "normalized" reciprocal space of a DWCNT, presented in Figure 2f[46]; this normalization consisting in the effective stretching or compression of DWCNT Brillouin zones is due to nanotube curvature. The coupling between states can be derived within the second-order perturbation theory in the following form[20,46]:

$$\delta E_{ii}^{el} = \sum_{\beta=1}^{3} \frac{|A_{i\beta}|^2}{\Delta E_{i\beta}}$$

It is determined by the term $A_{i\beta}$, *i.e.* the sum of the coupling matrix elements $M_{ii}$ for valence and conduction bands, and the term $\Delta E_{i\beta}$ representing the energy difference $\Delta E_{i\beta} = E_{ii} - E_{\beta}$ between coupled inner-tube ($E_{ii}$) and outer-tube ($E_{\beta}$) states; and the sum goes over three sites in the normalized Brillouin zones of a DWCNT (black dots in Figure 2f), which contain equivalent and non-equivalent states of inner and outer tubes, respectively. The $M_{ii}$ elements have a simple form and are mainly sensitive to interlayer distance $\Delta r$ (Figure 3a). The $\Delta E_{i\beta}$ term shows an intricate dependence on the inner and outer tube chiral angles for a fixed inner-wall tube diameter (Figure 3b) and a fixed optical transition (Figure 3c) and is largely responsible for the rich dependence of energy shifts on DWCNT geometry. Apart from this interaction, there is an additional constant redshift $\delta E^{diel}$ associated with the change of the effective dielectric screening in the presence of the adjacent tube[52]. It is estimated as −50 meV (−60 meV) for metallic (semiconducting) layers[20,46]. Summing up two contributions gives the total energy shift:

$$E_{ii}^{total} = \delta E_{ii}^{el} + \delta E^{diel}$$

Liu *et al.* and Zhao *et al.* showed experimentally that this total shift varies in DWCNTs from −200 to +50 meV[20,21] with respect to the transitions in isolated SWCNTs, and that the same inner layer can have different resonance energies depending on the geometry of the outer tube (Figure 3d).

Interestingly, the discussed intertube electronic coupling can become exceptionally strong in commensurate DWCNTs or those satisfying specific geometric conditions[53]. Most of these DWCNTs are challenging to observe experimentally due to low occurrence probability, so they have been studied so far only theoretically. Theory predicts significant modifications of the layers' electronic band structure, *e.g.*, semiconducting-to-metal transitions[54], the appearance of mini-Dirac points and pseudo-gaps[44]. For instance, Koshino *et al.* demonstrated the emergence of flatbands[53] in the (26,3)@(35,3) DWCNT by comparing the energy dispersion relations for uncoupled and coupled DWCNT layers (Figure 3e and Figure 3f, respectively).

The vdW interaction is also responsible for the emergence of new intertube electronic transitions, which were recently observed experimentally by Zhao *et al.*[22] and described theoretically by Rochal *et al.*[46] and Popov *et al.*[47] For example, Figure 3g shows a Rayleigh spectrum of an individual (12,11)@(17,16) DWCNT, with dashed lines indicating the positions of the optical resonances of the constituent SWCNTs[22]. The total number of the observed transitions in the DWCNT is larger than the sum of those from SWCNT resonances. It was predicted[46] that when the conduction band minima and valence band maxima shift in a certain way due to vdW coupling, the unconventional inter-tube electronic transitions become possible between van Hove singularities of inner and outer tubes (black arrows in Figure 3h).

The final aspect of interlayer electronic coupling consists of the significant modification of the excited state dynamics by creating fast non-radiative decay channels (Figure 3i). Femtosecond time-resolved luminescence measurements[55] and time-domain *ab initio* simulations[56] estimated the exciton energy transfer from the inner to the outer semiconducting tubes as fast as ~150 fs. A well-known consequence of such rapid non-radiative decay is the photoluminescence quenching of the inner layers of DWCNTs, which leads to at least four orders of magnitude decrease of the PL intensity relative to freestanding SWCNTs[24]. In principle, this quenching can be manipulated through 1D heterostructures engineering, *e.g.*, using isolating outer boron nitride layers instead of CNT, as will be discussed further in other sections of this review.

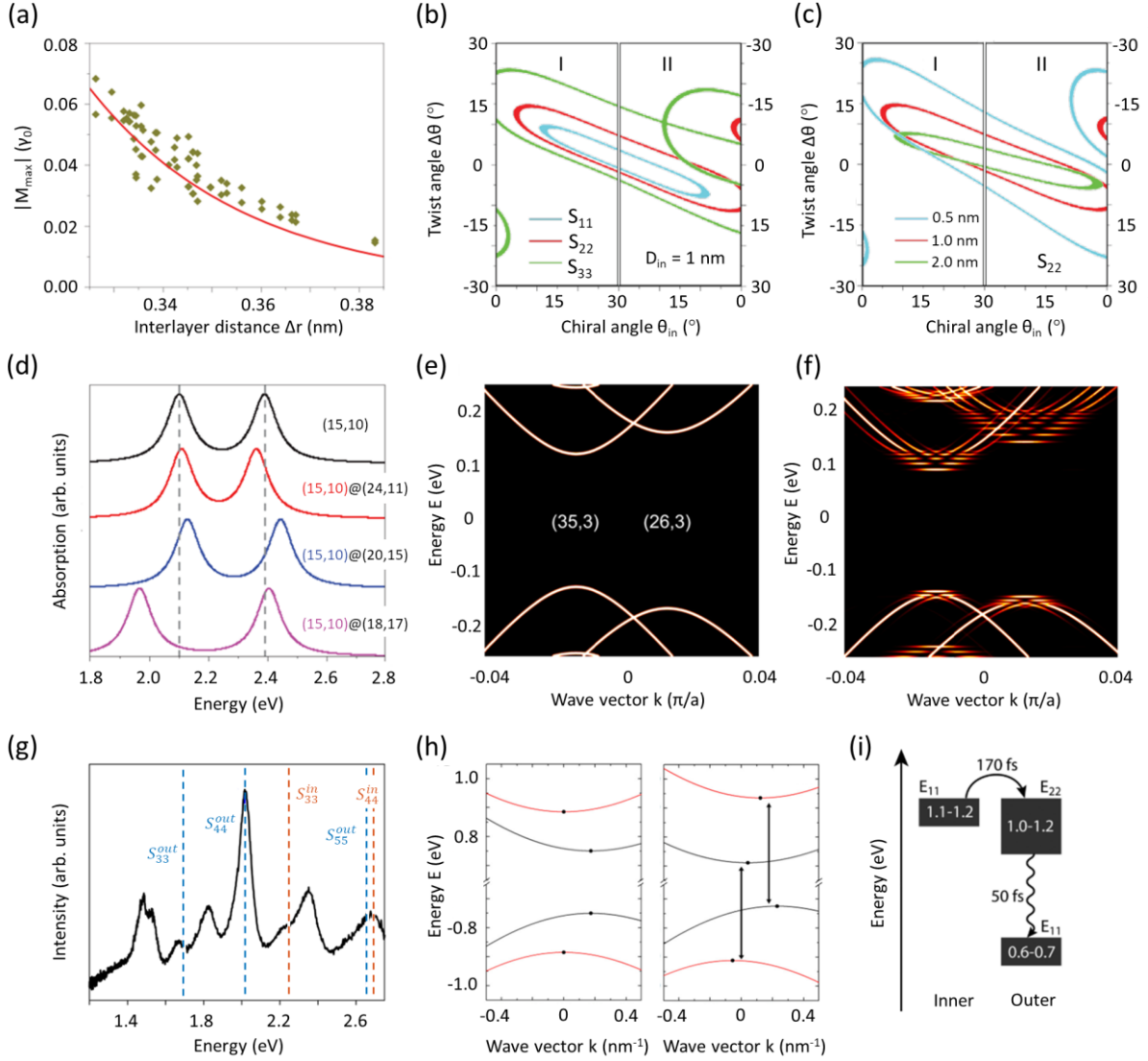

Figure 3. **Electronic van-der-Waals coupling in DWCNTs.** (a) Strong exponential dependence of the electronic coupling matrix element |$M_{max}$|, *i.e.* the maximum value of $M_{ii}$, on the interlayer distance $\Delta r$ (adapted from ref.[20] Copyright 2014, Nature Publishing Group.). (b) Dependence of the largest value of the |$1/\Delta E$| electronic coupling term (shown as contours) on the inner-tube chiral angle, twist angle and excited inner-tube $S_{ii}$ transition in DWCNTs with $D_{in}$ = 1.0 nm and $\Delta r$ = 0.35 nm (adapted from ref.[20] Copyright 2014, Nature Publishing Group.). (c) Dependence of the largest value of the |$1/\Delta E$| electronic coupling term (shown as contours) on the inner-tube chiral angle, twist angle and inner-tube diameter $D_{in}$ for the $S_{22}$ transition. The intertube spacing $\Delta r$ is fixed to 0.35 nm (adapted from ref.[20] Copyright 2014, Nature Publishing Group). (d) Simulation of the inter-tube electronic coupling-induced optical transition shift ($S_{33}$ and $S_{44}$) for a (15,10) inner SWCNT in different outer tubes (adapted from ref.[20] Copyright 2014, Nature Publishing Group). (e) Energy dispersion relations of decoupled (26,3) and (35,3) SWCNTs (adapted with permission from ref.[53] Copyright (2015) by the American Physical Society). (f) Energy dispersion relations of (26,3)@(35,3) DWCNT, revealing flat bands in conduction and valence energy bands (adapted with permission from ref.[53] Copyright (2015) by the American Physical Society). (g) The appearance of intertube optical transitions in the Rayleigh spectrum of a (12,11)@(17,16) DWCNT. Dashed red and blue lines indicate the expected optical transition energies for pristine inner and outer nanotubes, respectively (adapted with permission from ref.[22] Copyright (2015) by the American Physical

Society). (h) Energy dispersion relations of the decoupled (12,12) and (21,13) SWCNTs (left panel) and (12,12)@(21,13) DWCNT (right panel), revealing the condition for the appearance of intertube transitions (arrows) (adapted with permission from ref. [46] Copyright (2015) by the American Physical Society). (i) Schematic representation of an ultrafast exciton transfer between inner and outer layers of a DWCNT. Adapted with permission from [56]. Copyright 2015, American Chemical Society.

Finally, the VDW interaction can also affect the phonon properties of 1D heterostructures. Such coupling is often called mechanical and varies mainly with the interlayer distance[19,57] (changing by over two times over the range of inter-tube separation existing in DWCNTs, Figure 4a) and has negligible dependence on the twist angle[42]. Liu *et al.*[19] compared the vdW interaction strength in DWCNTs with the vdW coupling between unit-area graphene sheets under pressure obtained from compressibility measurements of graphite and showed that the effective pressure between the walls of as-grown DWCNTs reaches GPa (upper scale in Figure 4a). Later, Popov *et al.*[58] demonstrated that this effective pressure results from structural optimization of DWCNT layers, leading to radius changes (expansion/compression) as a function of both the diameter and the interlayer distance (Figure 4b). It is interesting to note that DWCNTs with $\Delta r > 0.34$ nm allow readily access a negative pressure on inner SWCNTs in the GPa range, which is difficult to achieve using conventional approaches.

In practice, the mechanical vdW coupling is manifested in the changed nature of Raman-active modes[23,59] (*e.g.*, the appearance of in-phase and out-of-phase vibrations as shown in the inset of Figure 4c), modulations of phonon frequencies[19,23,57] and the different resonance behavior[19,23]. The frequency change is illustrated in Figure 4c by the upshift of the observed low-frequency ($\omega_L$) and high-frequency ($\omega_H$) RBMs in DWCNT relative to the expected RBM frequencies of isolated SWCNTs (dashed lines). The changed resonance behavior is clear from Raman excitation profiles (Figure 4d), where the coupled RBMs have non-zero intensity both at the inner and outer tubes excitonic resonances (*e.g.*, at $M_{33L}$ and $S_{55}$ transitions in DWCNT (22,14)@(40,1), Figure 4d). Hence, one can observe the RBM of a specific layer even when its electronic transitions are not in resonance with the excitation laser[19,23]. Finally, the main consequence of the large effective pressure in DWCNTs is the G-mode splitting as a function of interlayer distance[60] (Figure 4e). The splitting is negligible in DWCNTs with $\Delta r \approx 0.335$ nm (top panel in Figure 4e) but grows significantly at large inter-tube separations (0.349 and 0.366 nm in the middle and bottom panels of Figure 4e, respectively).

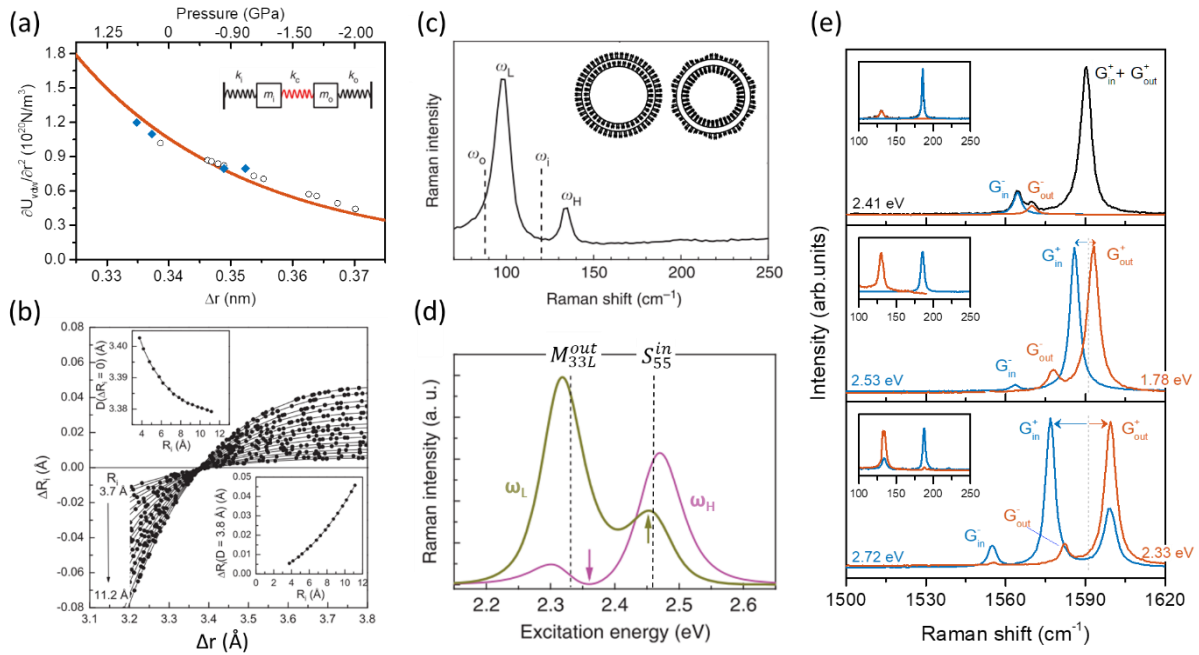

Figure 4. **Mechanical van-der-Waals coupling in DWCNTs.** (a) Dependence of the VDW unit-area force constant on the inter-tube distance $\Delta r$. The data points are derived from the experimental RBM frequencies in DWCNTs within the coupled oscillators model. The different inter-tube separation in DWCNTs corresponds to an effective pressure variation from 0.75 to -2.0 GPa (top label) (adapted with permission from ref. [57] Copyright 2017, Elsevier). (b) Calculated pressure-induced change of the inner-tube radius of DWCNTs, $R_i$, for DWNTs with given $R_i$ from 3.7 to 11.2 Å and different outer-tube radius $R_o$ (reprinted figure with permission from ref. [58] Copyright (2018) by the American Physical Society). (c) Raman spectrum for an individual (15,13)@(31,4) DWCNT, revealing the strong blue-shifted of DWCNT RBM frequencies from those of the constituent SWCNTs (dashed lines) (adapted from ref. [19], Copyright 2013, Nature Publishing Group); the inset shows collective in-phase (left) and out-of-phase (right) vibrations with frequencies $\omega_L$ and $\omega_H$, respectively, emerging due to mechanical coupling (reprinted figure with permission from ref. [59], Copyright (2002) by the American Physical Society). (d) Simulated Raman excitation profile of out-of-phase ($\omega_H$) and in-phase ($\omega_L$) RBMs for the (22,14)@(40,1) DWCNT with the quantum interference effect. REPs reveal the changed resonance conditions due to the mechanical coupling of two layers: contrary to the SWCNT case, the outer tube RBM $\omega_L$ (inner tube RBM $\omega_H$) can also be observed when exciting the inner tube $S_{55}$ transition (outer tube $M_{33L}$ transition) (adapted from ref. [19] Copyright 2013, Nature Publishing Group). (e) Experimental dependence of the G modes' frequencies in DWCNTs on the effective pressure and interlayer interaction. From top to bottom, G-modes and RBMs (inset) of three individual DWCNTs with increasing values of inter-layer distances $\Delta r$ (0.335, 0.349 and 0.366 nm). Reproduced with permission from ref. [60] Copyright 2017, Wiley-VCH.

## 3. SWCNT heterostructures by endohedral filling:
## 3.1 Molecular confinement and alignment inside SWCNTs

The confinement of molecules inside the 1D hollow core of SWCNTs can yield new properties for the encapsulated molecules and the host SWCNTs and create new functionalities due to the vdW interaction between the 'filler' and the 'host'.

An excellent example of the first case is the very peculiar phase behavior of water confined inside SWCNTs. Molecular dynamics simulations found that water molecules confined inside different SWCNT diameters adopt an arrangement with optimized hydrogen bonds, resulting in at least nine different ice phases with very different phase transition temperatures for diameters in the range of 0.9 - 1.7 nm.[61,62] Experimental verification of this diameter-dependent phase behavior was obtained recently through optical spectroscopy, using the vibrational[63] and electronic[64,65] properties of the SWCNTs as ultrasensitive probes to detect changes in the confined water structure. Indeed, both the radial breathing vibrational frequency of the SWCNTs[66,67] and their optical transitions[66,68,69] were found to shift and broaden after encapsulation of water, originating from an additional restoring force acting on the RBM vibration and a screening of excitons by the encapsulated dielectric. The observed shifts depend not only on the specific diameter and chiral structure of the SWCNTs[67], but also on the specific shape and structure of the encapsulated molecules[70] so that they can be used to determine for each filler molecule the minimal diameter in which it can be encapsulated.[70–72] Interestingly, when monitoring the emission of water-filled SWCNTs as a function of temperature, and comparing this with the emission from the same but empty SWCNTs, it is possible to observe phase transitions of water, confined inside the SWCNTs, through their effect on the optical transitions of the SWCNTs. For example, Chiashi *et al.*[65] probed a shift in the optical emission from individually suspended SWCNTs filled with water molecules , which was not observed for the empty analogues, at low temperature and thereby revealed a very complex behavior of the melting points of water inside SWCNTs with different diameters, in agreement with the theoretical predictions. Interestingly, besides these freezing transitions of liquid water inside broader-diameter SWCNTs, other quasi-phase transitions could be observed for very small diameters ($d <$ 0.9nm), in which only a single file of water molecules can be encapsulated.[67,73] In this case, at low temperatures, the orientations of the water dipoles all align in the same sense along the SWCNT axis (*i.e.,* forming a ferro-electric chain of water dipoles), resulting in a large internal dipole that shifts the emission of the surrounding SWCNTs.[64] Although interesting results were obtained, a large range of SWCNT diameters still remains to be investigated, and only the behavior of confined water has been elucidated so far, leaving a legion of solvent/SWCNT combinations[70] to be explored.

The filling of SWCNTs can also lead to a change of the SWCNT properties. First of all, as mentioned above, SWCNTs can be p- or n-doped, with the doping level depending on the electron affinity or ionization potential of the encapsulated molecules[30]. By diluting the donor/acceptor molecules with similar-shaped inert molecules, and thereby creating alternating patterns of filling, the doping level can be tuned, as recently demonstrated,[74] enabling the already unique electronic properties of SWCNTs to be further enhanced for potential applications in nanoelectronics.[74] Secondly, because of the presence of the encapsulated molecules, the mechanical properties of filled SWCNTs are altered, typically resulting in higher stability under the application of pressure,[75] similarly as observed for DWCNTs.[76] Most importantly, the optical transition energies at which SWCNTs absorb and emit light can be tuned by filling SWCNTs with solvents with varying dielectric constants.[70] Surprisingly, even for those diameters in which these solvents can just fit with a single or double file of molecules, the optical transitions energies of the SWCNTs still depend on the dielectric constant of the bulk solvent, and stacking-dependent effects only represent a perturbation on this general dielectric constant trend.[70]

Encapsulation of solvent molecules, like water, typically occurs spontaneously after opening one of the SWCNT ends.[77] Even a very short (30s) exposure of opened SWCNTs to the solvent can lead to a complete filling of the SWCNT's hollow core,[70] indicative of a swift, collective motion of the solvent molecules into the hollow interior.[78] Opening of SWCNTs typically occurs during the purification procedure, but can also be achieved by simple sonication.[66] For organic molecules, various methodologies for filling have been reported. While filling through gas-phase sublimation can only be used for those molecules stable under such conditions,[30,79] refluxing in a saturated solution of the molecules in a solvent[71,72,80,81] or nano extraction from supercritical $CO_2$[82,83] can in principle be employed for any molecules. However, the former has the risk of also encapsulating the solvent, while the latter is most effective for molecules with a low affinity for supercritical $CO_2$.[84] Therefore, for each filler molecule, the encapsulation methodology can be selected based on its solubility in different solvents, its stability at high temperatures and its affinity for supercritical $CO_2$. Afterwards, externally adsorbed molecules must be carefully removed, which can be done either chemically, *e.g.*, by soaking the SWCNTs in a piranha solution,[85] or physically, by excessive rinsing of the SWCNTs with a solvent in which the molecule readily dissolves.[71,72,80,81,86] Important for all these procedures is the preparation of reference samples to unambiguously distinguish filling from external adsorption, *e.g.,* performing the same procedure with closed SWCNTs or exposing the SWCNTs separately to each of the other solvents they were in contact with.[71,72]

Most interesting is the exploitation of the 1D SWCNT hollow structure as a template to create new functionalities. The SWCNT diameter is an essential parameter in determining the arrangement of the encapsulated materials. When thinking of atoms and fullerenes confined inside CNTs, the arrangement of these nanospheres inside the cylindrically confined space can lead to different ordered phases, which were first calculated for hard spheres by Pickett *et al.*,[87] and later confirmed for encapsulated fullerenes (Figure 5a).[88] Also, stable atomic metal nanowires can be obtained by encapsulation and confinement inside thin-diameter SWCNTs.[26,89,90] Such 1D crystals are anticipated to have very peculiar transport, electronic and magnetic properties as opposed to their bulk analogues,[91] with the CNT serving both as a template, to isolate the atomic chains, and as a protection that stabilizes the chains, so that they can be studied avoiding oxidation taking place. Besides, the synthesis and subsequent isolation of 1D transition metal monochalcogenides remained elusive until they were confined inside SWCNT templates (Figure 5b).[92] The simple synthesis through vacuum annealing of $MoTe_2$ bulk crystals in the presence of opened CNTs gives rise to a high-yield, bottom-up assembly of these isolated MoTe wires and raises opportunities to explore also the physics of these 1D heterostructures.

Most filling-related research, however, has been devoted to the encapsulation of organic (dye) molecules inside the SWCNTs, where filler-host interaction can lead to a photosensitization of the SWCNTs through energy or charge transfer from the encapsulated molecules to the SWCNTs.[72,79,81,93,94] Encapsulation of dyes inside SWCNTs can furthermore be used to create heterostructures with unprecedented Raman cross-sections. This is due to a combined effect of confining the dye molecules inside the SWCNTs, the formation of highly polarizable J-aggregates and the fact that fluorescence of the dyes is fully quenched by the surrounding CNT such that these heterostructures are free from any fluorescence background.[85] Moreover, the increased stability of the dyes with respect to photobleaching[80] and the possibility to modify the SWCNT surface covalently towards biological compatibility without altering the Raman signals from the encapsulated dyes led to the formation of heterostructures for Raman imaging of biological systems.[85] Koyama *et al.* found by TEM that, also in this case, the diameter of the SWCNTs is crucial in obtaining either a single or double-file of encapsulated molecules (Figure 5c).[95] For a single-file case, they observed the molecules to be localized off-center due to the vdW stacking with the SWCNT walls and demonstrated that the excitation energy transfer from molecule to the SWCNTs depends on the formation of these single- or double-file arrays. In a later study, they surprisingly found that not only the diameter but also the electronic character of the SWCNTs, *i.e.*, metallic or semiconducting, would determine the specific stacking of the molecules inside their hollow core, which was explained in terms of the vdW interaction between the encapsulated molecules on the one hand and between the molecules and the SWCNT walls on the other hand.[96]

A diameter-dependent stacking of dye molecules inside SWCNTs with different diameters was also found by S. van Bezouw *et al.*[72], who observed a new peak arising in photoluminescence excitation (PLE) spectroscopy maps. It originated from the light absorption by the encapsulated dyes and the subsequent emission from the SWCNTs via an efficient energy transfer between the filler and the host (Figure 5d). By carefully fitting the PLE data of multiple samples with different diameter distributions, the absorption of the dyes was found to depend strongly on the diameter of the host SWCNT and was attributed to different dye stacking geometries in SWCNTs with different diameters.

Aside from the above-mentioned diameter and electronic character effect, E. Gaufrès *et al.* demonstrated that the synthesis conditions could also result in different configurations in the dye stacking.[86] By preparing numerous samples through the solvent-based filling methodology using different synthesis conditions, they found that either single- or double-file arrangements could be obtained depending on the dye concentration and the synthesis temperature (see Figure 5e).[86] A mechanism was proposed in which single-file filling occurs spontaneously (even at low concentration and temperature), while the pair-aggregate filling was found to be endothermic. These novel insights that the exact filling procedure can dramatically alter the arrangement of dye molecules inside CNTs and the resulting properties of the system are fundamental to the development of such 1D heterostructures. Thus, aside from the diameter of the SWCNT and the above-mentioned concentration and temperature in the solution-based filling, it can be expected that other parameters can strongly influence the arrangement of encapsulated molecules, *e.g.* the affinity of the dye to the solvent or supercritical $CO_2$ and the intermolecular interactions between the dyes themselves.

The specific mechanism of the efficient filling of SWCNTs with many molecules is still unclear. Depending on the confinement effects (size of molecule versus CNT diameter) and vdW interactions of the dyes with each other and with the SWCNT wall, the molecular motions are expected to be restrained which should influence the encapsulation processes. A good geometric fit between the molecules and the SWCNT walls could even further stabilize the vdW interactions, resulting in the irreversible filling of the SWCNTs with the various compounds. On the other hand, flow of water through SWCNTs is extremely high[97], orders of magnitude higher than predicted by macroscopic diffusion models, and also for larger dye molecules it was found that SWCNTs with lengths up to 10 μm were readily filled with the dyes, indicating that once inside, the molecules easily move across these long lengths and fill the endohedral space completely.[86]

Another excellent example in this respect is the alignment of dipolar dye molecules inside the hollow core of SWCNTs. For sufficiently small SWCNT diameters, the dipole-dipole interaction between the dyes should automatically create a 1D head-to-tail arrangement of these asymmetric molecules, with all electric dipoles pointing in the same sense. Figure 5f presents a model of such a configuration that is highly desirable for second-order nonlinear optics and which cannot be spontaneously achieved in any higher dimensions.[71] The first experimental demonstration of this alignment found that at least 70 molecules could be aligned inside the hollow core of the SWCNTs. It resulted in a heterostructure with a giant static second-order polarizability due to the coherent addition of head-to-tail ordered arrays of those 70 molecules.

One of the main observations for encapsulated dyes inside SWCNTs is that their emission is strongly quenched after encapsulation due to the efficient energy transfer from the dyes to the lower-band-gap SWCNTs[71,72,80,81,85,86]. It is similar to the quenching of the inner tube fluorescence in DWCNTs, where the larger diameter outer CNTs typically have a smaller band gap than the inner tubes.[24] The PL quenching is nearly complete, *i.e.* quenching efficiencies of up to 99.999% have been found for encapsulated species[85]. The lower reported PL quenching values[95,98] were most likely originating from trace amounts of externally adsorbed molecules still present in the samples. This exemplifies the importance of proper reference samples to assess the efficiency of removing any externally adsorbed dyes from the sample, *e.g.*, a sample with closed SWCNTs treated in the exact same way as the opened SWCNTs.[72]

The PL quenching can be advantageous when studying the nonlinear optical properties of the heterostructures[71] or their Raman cross-sections[85]. However, excluding the most crucial property of organic dyes, *i.e.,* emitting light at various wavelengths, can be a substantial disadvantage, in particular when thinking of bioimaging applications. To solve this issue, very recently, boron nitride nanotubes (BNNTs) have been similarly filled with dye molecules.[99] The much larger bandgap of BNNTs and their corresponding optical transparency in the visible and IR provide the ideal homogeneous environment for the confined dyes, protecting them from photochemical damage while maintaining and even improving the strong emission from the dyes. Indeed, it was demonstrated that the stability of the dyes in terms of bleaching and blinking was increased by more than 4 orders of magnitude after encapsulation in BNNTs.[99] The aggregation inside the confined BNNT interior moreover resulted in a shifted dye emission, even towards the second optical transparency window in the near-infrared, ideally suited for bioimaging applications.

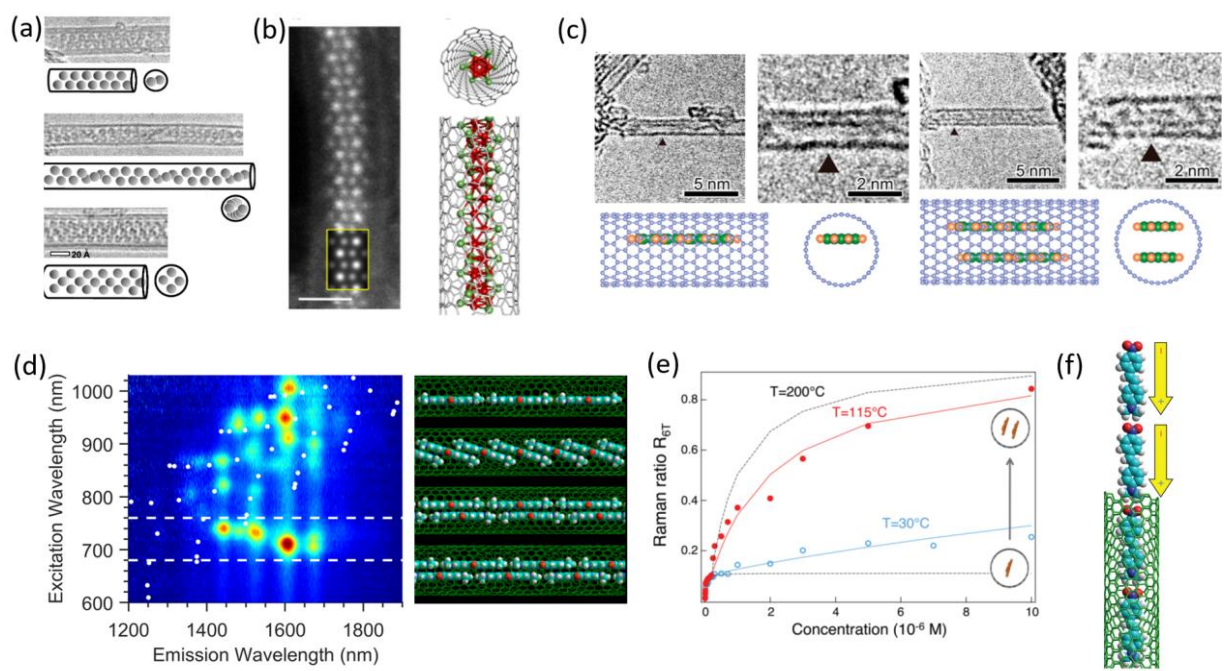

Figure 5. **Stacking of atoms/molecules inside SWCNTs with different diameters**. (a) Different ordered phases of fullerene stacked inside SWCNTs with different diameters.[88] (b) Arrangement of MoTe nanowires inside carbon nanotubes.[92] Reprinted with permission from [[92]]. Copyright 2019 American Chemical Society. (c) TEM images and corresponding schematic representation of quateryllene molecules encapsulated inside thin (left) and thick (right) diameter SWCNTs.[95] Reprinted with permission from [[95]]. Copyright 2014 American Chemical Society. (d) PLE map of squaryllium dyes (1,3-bis[4-(dimethylamino)phenyl]-squaraine) inside SWCNTs with various diameters.[72] White points indicate SWCNT peak positions, while the dashed lines highlight the position of additional EET peaks, *i.e.* absorption of the dye followed by energy transfer to the SWCNTs and then emission from the SWCNTs. The absorption wavelength of the squaryllium dyes changes with surrounding SWCNT diameter, indicative of different stacking geometries inside the SWCNTs. A model describing possible configurations of squaryllium dyes inside SWCNTs with different diameters, resulting in different aggregate-couplings, which on their turn result in different absorption wavelengths for the encapsulated dyes. Reprinted with permission from [[72]]. Copyright 2018 American Chemical Society. (e) Raman ratio, *i.e.* Raman intensity of encapsulated α-sexithiophene molecules with respect to the SWCNT G-band at 532nm, as a function of dye concentration in toluene after the synthesis at two different temperatures (30°c and 115°C),[86] showing that concentration and temperature define the formation of single or double files of encapsulated α-sexithiophene molecules; Reprinted with permission from [[86]]. Copyright 2016 American Chemical Society. (f) head-to-tail dipolar alignment of p,p'-dimethylaminonitrostilbene (DANS), *i.e.* a dipolar dye molecule, inside SWCNTs with sufficiently small diameters. A theoretical model of these molecules in a (9,7) SWCNTs is presented, which was found to be the smallest SWCNT diameter in which the dye can be encapsulated.[71] Reprinted with permission from [[71]]. Copyright 2015, Nature Publishing Group.

## 3.2 Transformation of precursor molecules inside SWCNTs

The 1D hollow structure of SWCNTs also provides the smallest nanoreactor for the templated synthesis of various compounds. The simplest example in this respect is the thermal conversion of fullerenes into DWCNTs, through the thermally-activated coalescence of the fullerenes inside an 'outer' SWCNT.[100–102] By controlling the initial SWCNT diameter and the specific precursor used, very thin inner tubes can be synthesized that are otherwise very difficult to achieve.[103] Isotopically-hetero DWCNTs (*i.e.*, SW$^{13}$CNT@SW$^{12}$CNT) were also synthesized by this approach.[104] It was demonstrated that thermal conversion of $C_{60}$ mainly grows (6,5)/(6,4) inner tubes, while thermal conversion of $C_{70}$ preferentially forms (6,4) inner tubes. The conversion of ferrocene yields (6,5)/(9,1) tubes and the coalescence of a perylene-derivative (perylene-3,4,9,10- tetracarboxylic dianhydride or PTCDA) preferentially leads to (8,1)/(7,2) inner tubes.[103]

The reaction path proposed for fullerenes consists of their coalescence followed by a thermal annealing to repair most defect sites and thus synthesize defect-free inner DWCNT layers (see Figure 6a). Similarly, small precursors like pentacene rotate freely to merge into deformed fullerene-like intermediates that afterwards coalesce into inner DWCNT layers. In contrast, the fusion of larger molecules like PTCDA results first in a polymerization towards graphene nanoribbons (GNRs) that then twist and fuse together to form an inner tube of a DWCNT, thereby resulting in other chiralities to be grown.[103] Chamberlain *et al.* verified this synthesis of nanoribbons as an intermediate step to the formation of DWCNTs, through time-dependent HRTEM imaging starting from SWCNTs filled with perchlorocoronene (PCC, see Figure 6b). After filling, these PCC molecules form 1D stacks, separated by the optimum vdW distance, inside the SWCNTs due to the strong intermolecular π-π-interactions (top left panel in Figure 6b). When increasing the electron dose level, the intermolecular interactions are broken from the end of the stack to the middle, tilting the molecules within the tubes such that they can be polymerized into planar GNRs. Further increasing the electron dose twists these graphene nanoribbons, as presented in the last panel of Figure 6b.[105] Similar observations could be made by Raman spectroscopy, where each of the intermediate stages can be followed through their very distinctive Raman responses. Figure 6c provides such an example: Raman spectra of SWCNTs filled with coronene (in blue) using two different filling methods (extraction from supercritical $CO_2$ (SC, top panel) and low-temperature gas-phase sublimation (LT, bottom panel)) are compared to unfilled SWCNTs (in black) and to Raman spectra of filled SWCNTs after annealing at different temperatures ranging from 500-1250°C.[83] In the range of 500-900°C (green and yellow spectra), new Raman features around 1200-1400cm$^{-1}$ are indicative for the formation of GNRs, while after further annealing to 1250°C (red spectra), new features in the low-frequency regime are observed, indicative of the growth of inner SWCNTs.[83]

Similarly to SWCNTs, BNNTs can be filled with fullerenes that can be further transformed into an inner SWCNT by thermal annealing, resulting in SWCNT@BNNT heterostructures.[106] Unlike in CNTs, the fullerenes show weak host-guest interactions and can be easily removed by sonication from the host BNNTs. The smaller vdW interaction between SWCNT and BNNT may help in extracting the thin inner SWCNTs afterwards. Likewise, BNNT@SWCNT combinations can be obtained by filling SWCNTs with ammonia borane complexes followed by thermal transformation.[107] To achieve this synthesis, SWCNTs were end-capped with fullerenes that served as corks with high vdW interaction with the SWCNT walls, preventing volatile ammonia borane complexes from exiting the SWCNTs during the thermal transformation.[107] The significant achievement of this templated synthesis was the growth for the first time of thin diameter single-wall BNNTs with large band gaps (6eV) similar to the multi-wall BNNTs.

Besides the controlled formation of inner BNNT or CNTs inside SWCNTs, also the intermediate templated growth of GNRs is certainly highly attractive. Recently, it was demonstrated that by carefully controlling the starting SWCNT diameters and the exact annealing temperature, 6- and 7 armchair graphene nanoribbons (6- and 7-AGNRs) with well-defined structure were grown inside SWCNTs, and their electronic and vibrational properties were characterized in detail by wavelength-dependent Raman scattering spectroscopy (Figure 6d).[108] The Raman spectra show the features of the surrounding SWCNTs (RBMs, D and G-bands), those of newly synthesized inner tubes in the RBM region (highlighted by the circle in Figure 6d) and the distinctive radial-breathing-like-modes (RBLM) and higher-frequency modes of the encapsulated AGNRs (highlighted by the squared boxes in Figure 6d). Interestingly, the synthesis of GNRs inside SWCNTs can be achieved both from functionalized fullerenes, polycyclic aromatic hydrocarbons, and ferrocene, thus showing that mainly carbon-rich precursors are needed combined with other atoms (H, O, N, S, …) to terminate the edges.[109–111] Since both the GNRs and the CNTs can be metallic or semiconducting, depending on their structure, a great diversity of heterostructures with different electronic and optical properties can be synthesized with this method. Moreover, the templated synthesis allows growing very thin nanoribbons, which are otherwise very hard to achieve.

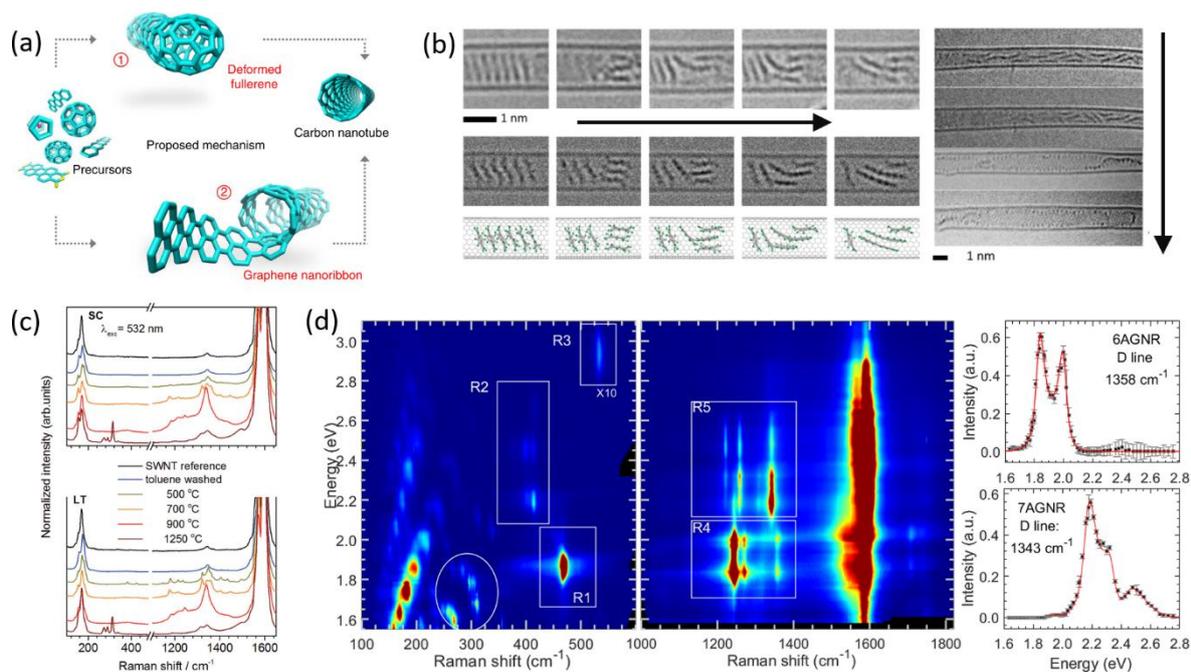

Figure 6. **Transformation of precursor molecules into graphene nanoribbons and inner SWCNTs**. (a) Proposed reaction paths for the thermal conversion of various precursors into well-defined inner tubes of DWCNTs, that is, (1) the conventional coalescence of fullerenes and (2) the formation of graphene nanoribbons that twist to form a DWCNT.[103] Reprinted with permission from [[103]]. Copyright 2013, American Chemical Society (b) Transformation of encapsulated perchlorocoronene (PCC) molecules from 1D stacks to GNRs followed by HRTEM by increasing the dose level of the electron beam (black arrow pointing in the direction of higher dose levels).[105] Initially, encapsulated PCC molecules form vdW-stacked ordered arrays inside the SWCNTs (upper left image and simulation). As the dose increases, intermolecular vdW interactions are broken from the end towards the middle of the stack, followed by a planarization and at the highest dose levels a GNR is formed, with the Cl-atoms at the edges providing the necessary contrast. This picture was derived with permission from [[105]], Copyright 2017, American Chemical Society (c) Raman spectra obtained at an excitation wavelength of 532nm of coronene-filled SWCNTs through either a low-temperature gas-phase sublimation (LT, bottom panel) or through supercritical $CO_2$ filling (SC, top panel), transformed into GNRs at lower temperatures and inner SWCNTs at higher temperatures by thermal annealing.[83] Reprinted with permission from [[83]], Copyright 2013, Wiley VCH. (d) Wavelength-dependent Raman characterization of GNRs inside SWCNTs. Aside from the radial breathing modes of the SWCNTs (100-250cm$^{-1}$) and the G-band of the SWCNTs (1600cm$^{-1}$), signatures from the growth of inner SWCNTs (circled area) as well as 6 and 7 AGNRs (squared regions) can be observed. Regions R1 and R4 correspond to the RBLM and high-frequency modes of the 6 AGNRs, regions R2 and R5 to those of the 7 AGNRs. The right panels present a selection of the electronic resonance profiles for the D-modes of the AGNRs, plotting Raman intensity as a function of excitation energy, directly yielding information on the electronic structure of the AGNRs.[108] Adapted with permission from [[108]], Copyright 2021, Elsevier:

Besides the templated synthesis of GNRs and inner SWCNTs, the 1D nature of SWCNT templates can be used to create linear-chain nanodiamonds, by thermal annealing of specific precursor molecules,[112] or MoTe nanowires by thermal conversion of $MoTe_2$ nanocrystals in the proximity of opened SWCNTs. Again, the wide variability of precursor molecules that can be encapsulated in the wide variability of SWCNT diameters allows for designing a legion of new heterostructures, each with very different optical and electronic properties.

**3.3 Synthesis of linear carbon chains inside SWCNTs**

As a final example, linear carbon chains (LCC) are excellent examples of endohedrally synthesized nanostructures. Their synthesis in bulk has been limited to conjugated polyynes with up to 44 acetylenic carbons, as demonstrated by Chalifoux *et al.* in 2010[113]. The obstacle to create longer chains is the unavailability of longer and stable precursor molecules. Thin-diameter SWCNTs could provide an answer to form such linear chains, where the role of the SWCNT is to confine the carbon atoms to a single chain, as was similarly done for other atoms such as iodine.[114] The first evidence of the formation of LCCs inside MWCNTs was obtained by a combination of HRTEM and Raman spectroscopy in 2003, showing that much longer chains, up to at least 100 carbon atoms, could be synthesized.[115] A major breakthrough in this direction came in 2016, when L. Shi and co-authors demonstrated the synthesis of ultralong LCCs with up to 6000 contiguous carbon atoms in the inner hollows space of DWCNTs.[116] These ultralong LCCs can be nicely imaged in HRTEM (Figure 7a).[117] In this case, LCCs were synthesized by high temperature and high vacuum annealing of DWCNTs, with an optimal synthesis temperature of 1460°C. The DWCNTs provided both stability at high temperatures and also the required thin diameter since density functional theory (DFT) calculations predicted an optimal synthesis diameter of 0.69 nm, corresponding to the (5,5) chirality.[116] The LCCs show a single characteristic vibrational mode in Raman spectroscopy, the so-called C-mode, with extremely high Raman cross-section of the order of $10^{-22}$ cm$^2$ sr$^{-1}$ per atom, the strongest ever observed.[118] This C-mode originates from the in-phase stretching of the triple bonds. Due to the Peierls distortion, the chains are polyyinic with alternating single and triple bonds, of which the bond lengths are determined by the so-called bond-length-alteration (BLA), *i.e.*, the difference between the bond length between adjacent atoms. The BLA not only determines the vibrational frequency of the C-mode, but also determines the electronic (optical) properties of the LCCs. For short chains, the BLA is highly sensitive to the length and end-groups of the chains, the latter used for their stabilization.[113] However, when an ultralong LCC is synthesized inside the DWCNTs, its C-mode frequency does not seem to depend on length anymore (see the right panel of Figure 7b), but more on the surrounding inner SWCNT chirality. Indeed, a correlation is observed between SWCNT diameter and C-mode frequency (as presented in the middle panel of Figure 7b), where higher C-mode frequencies are found for LCCs encapsulated in larger diameter inner SWCNTs, which indicates that mainly the vdW interactions between the LCCs and surrounding DWCNTs determine the BLA of the chains. Besides, this variable BLA also results in largely changing optical bandgaps (ranging between 1.8-2.25eV), as verified by wavelength-dependent Raman spectroscopy in an ensemble sample containing many different DWCNT chiralities and thus distinct C-mode frequencies (Figure 7c)[117] as well as on the single tube/single chain level (Figure 7d)[119]. These experiments revealed that the SWCNT inner diameter could be an additional handle to tune the optical and vibrational properties of the encapsulated chains.

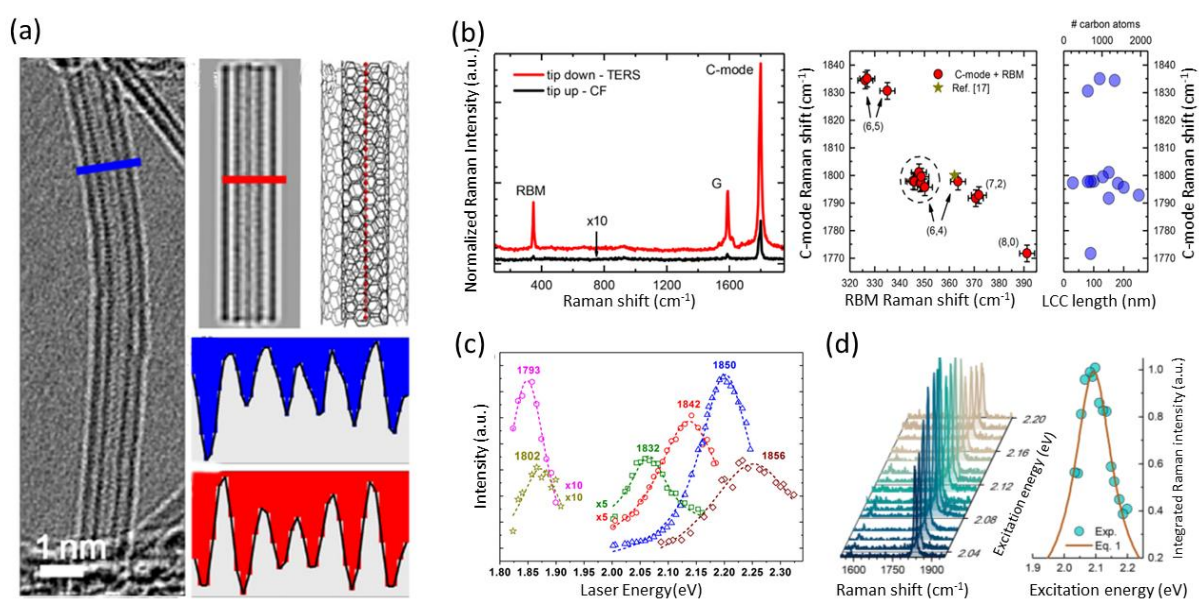

Figure 7. **Synthesis of linear carbon chains inside DWCNTs**. (a) HRTEM picture of ultralong linear carbon chains synthesized within DWCNTs, together with a TEM simulation and molecular model. Adapted with permission from [[117]], Copyright (2017) American Physical Society (b) Raman spectrum of an individual chain@DWCNT, with the RBM and G-band of the DWCNT combined with the C-mode of the encapsulated chain. The signal-intensity can be strongly enhanced by tip-enhanced Raman spectroscopy, also able to spatially map the length of the chains. The middle panel shows the frequency of the C-mode of the encapsulated chain with respect to the Radial Breathing Mode of the surrounding SWCNT, measured simultaneously with TERS. The Length of the chains, also obtained from TERS, is also compared. From this, it is clear that the C-mode frequency does not depend on the length of the chain, indicative of very long chains, but depends strongly on the confinement imposed by the surrounding nanotube.[120] Adapted with permission from [[120]]. Copyright 2018 American Chemical Society (c) Optical bandgap of encapsulated chains encapsulated inside different DWCNT chiralities (each color represents a different C-mode frequency, as indicated in the same color in the graph).[117] Reprinted with permission from [[117]], Copyright (2017) American Physical Society (d) Raman excitation profile of an individual encapsulated chain which can be measured due to the extremely high Raman cross-section of the encapsulated chains.[119] Reprinted with permission from [[119]], Copyright (2018) Elsevier

## 4. SWCNT heterostructures by external adsorption of molecules

Whereas the interior of CNTs can serve an ultrasmall space where unique behaviors in the filling, alignment, and chemical reaction of molecules are observed, the outer surface of the CNT can also be seen as an extremely curved substrate without dangling bonds. Early research on external SWCNT heterostructures was driven by the need for dispersing and isolating hydrophobic SWCNTs in water or other solvents; otherwise, as-grown SWCNTs form entangled bundles in which optical and electronic properties of the SWCNTs are greatly affected by the strong vdW interaction between the wide variety of SWCNTs. Among various functionalization techniques for SWCNT dispersion, physical functionalization through external adsorption on the SWCNT walls offers opportunities to tune interfacial properties of nanotubes without degrading their superior properties as opposed to chemical functionalization that introduces defects on the SWCNT walls.[121,122] Observation of fluorescence from isolated SWCNTs via sodium dodecyl sulfate surfactant[31] was a direct consequence of such functionalization and has brought excitement in the research community. It was found that in particular bile salt surfactants created ordered micellar structures around the SWCNTs, resulting in enhanced spectral resolutions in ensemble optical spectroscopy.[34] Furthermore, the coating of SWCNTs with single-stranded DNA sequences not only enabled the isolation of SWCNTs in water, but moreover resulted in a chiral structure recognition by the ordered barrel structure of the DNA sequences wrapping around particular ($n,m$) species, allowing for highly efficient chirality separation (Figure 8a,b).[32] Similarly, specific polymers such as poly(9,9-dioctylfluorenyl2,7-diyl) can selectively wrap and isolate specific SWCNT chiralities during solubilization.[33]

Another functionality of this wrapping is that the additional layer can protect the SWCNTs from interacting with their surrounding environment. Although such organic materials usually cannot provide better mechanical and chemical stability than SWCNTs themselves, they can prevent the typical nonradiative recombination of excitons in SWCNTs due to interactions with the environment.[123] Polystyrene, for example, is known to keep the fluorescence of SWCNTs bright when deposited on a substrate.[124] Core-shell structures of SWCNTs and polystyrene, where the SWCNTs were wrapped with a cross-linked polystyrene layer, therefore exhibited much brighter photoluminescence even on solid substrates (Figure 8c-e).[125] Such heterostructures may enable the use of SWCNTs as bright light sources that can be integrated into silicon-based photonic circuits and other photonic devices on solid substrates.

Although it is somewhat counter-intuitive, films of polymer-wrapped SWCNTs exhibit even better electrical transport properties than those with less polymers because isolation of a SWCNT from other SWCNTs leads to higher intra-nanotube mobility.[126,127]

In addition to the role of organic wrapping as protection, electronic and optical properties of SWCNTs can be modulated through charge or energy-transfer,[128–130] weak mechanical interactions,[131] and dielectric screening of excitons.[33,132] For example, when porphyrin oligomers stack through vdW interactions on SWCNT surfaces (Figure 8f,g), efficient energy transfer from the porphyrin to the host SWCNTs was observed,[133,134] combined with shifts of the optical transition energies that depend on the specific SWCNT chirality (Figure 8h).[131] The adsorption of various molecules on the SWCNT walls could therefore also be used as an effective tuning tool for the optical properties of SWCNTs, similarly as observed for endohedral filling[70], in particular for air-suspended SWCNTs, in which excitons are not strongly screened.[132] This implies that coating of SWCNTs either with locally adsorbed molecules or with an alternating pattern of molecules could result in tailored energy landscapes which could trap excitons and preserve bright fluorescence, allowing SWCNTs to function as quantum light sources, similarly as in the well-studied sp$^3$-functionalization, but without an actual chemical bond between the molecules and the SWCNTs.[122,124]

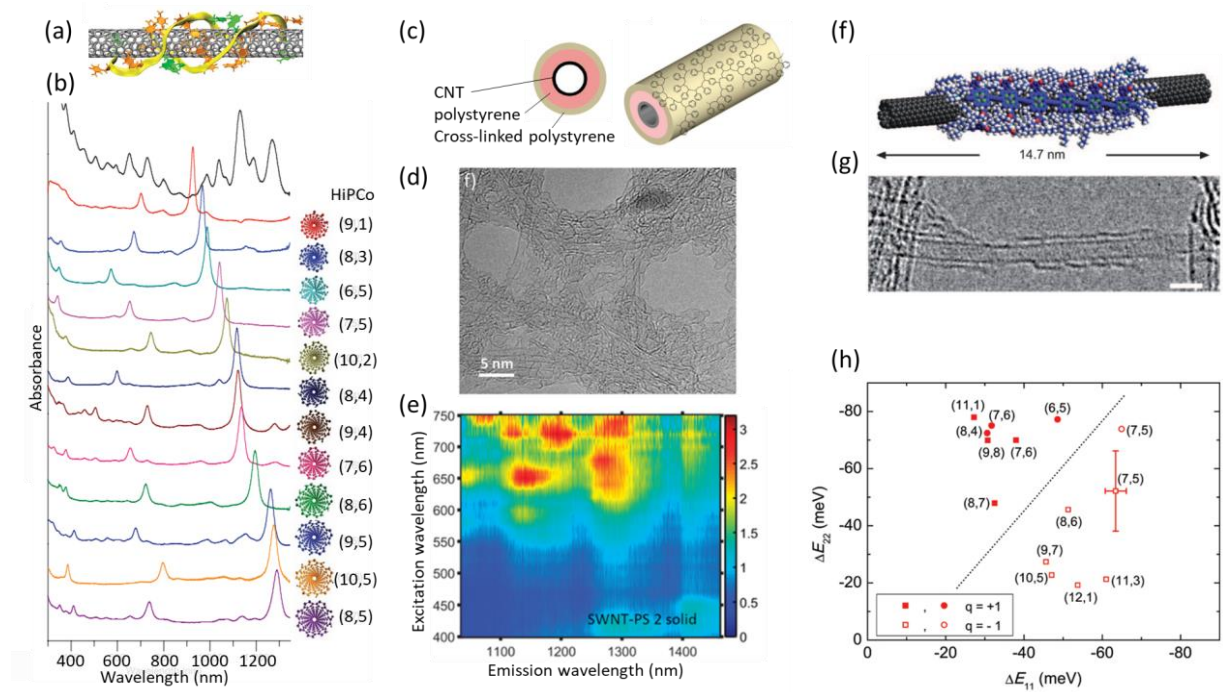

Figure 8. **One dimensional van der Waals heterostructures by organic wrapping.** (a) Schematic of DNA barrel on a (8,4) nanotube, composed of two hydrogen-bonded anti-parallel ATTTATTTATTT strands. (b) Absorption spectra of the starting SWCNT mixture (in black)

and 12 chirality-sorted semiconducting SWCNTs by using different DNA sequences. Panels (a,b) are reproduced with permission from [[32]]. Copyright 2007 Nature Publishing Group. (c) Schematic for core-shell functionalization of SWCNT by introducing polystyrene on the side walls, followed the formation of cross-linked polystyrene. (d) TEM images of SWCNT-polystyrene heterostructures. (e) PLE maps of a film of SWCNT-polystyrene on a polytetrafluoroethylene membrane. Panels (c-e) are reproduced with permission from [[125]]. Copyright 2018 Royal Society of Chemistry. (f) Molecular mechanics optimized structure of porphyrin hexamers and a (8,6) SWCNT. (g) HRTEM image of a heterostructure of SWCNTs and porphyrin hexamers. Panels (f,g) are reproduced with permission from [[133]]. Copyright 2011 Wiley-VCH. (h) Red-shifts in $E_{11}$ and $E_{22}$ transition energies relative to those of SDS-coated SWCNTs (*i.e.* the Weisman values)[[135]] as determined from PLE maps of SWNT-porphyrin heterostructures. Panel (h) is reproduced with permission from [[131]]. Copyright 2011 American Chemical Society.

## 5. Heterostructures of carbon nanotubes coated by hexagonal boron nitride

The first highly crystalized coating of SWCNTs was h-BN. In the early days of nanotube research, Suenaga *et al*. reported the formation and observation of coaxial heterostructures of CNTs and BNNTs by arc-discharge synthesis (Figure 9a).[[36]] CNT@BNNT heterostructures can most likely be easily formed because both materials consist of a one-atom-thick hexagonal lattice with very similar lattice constants (only 1.8% difference).[[136]] The recent surge in 2D materials research, where vdW heterostructures of graphene and 2D h-BN often play key roles,[[136–138]] has brought their 1D analogues back into the spotlight. In 2020, Xiang *et al*. reported 1D vdW heterostructures of SWCNTs and a few layers of h-BN with high crystallinity using a non-catalytic CVD process (Figure 9b,c).[[3]].

Since h-BN is a wide-band gap insulator, free of dangling bonds and with minimal charged impurities,[[139,140]] the effects of a surrounding h-BN layer around the SWCNTs on the electronic and optical properties of the SWCNTs are expected to be small in a similar manner as observed for their 2D analogues, such that the additional h-BN layer can serve as a homogeneous unperturbing and protecting environment, which in principle maximizes the potential of such CNTs in devices. The large bandgap of h-BN and the absence of prominent electronic coupling,[[141]] however, makes it difficult to identify the structures (*e.g.*, number of layers) and prove the purity and crystallinity of the h-BN layers by optical spectroscopy. One possible fingerprint of h-BN coating on CNTs is a downshift of the G-band from the CNTs, most likely originating from a weak interlayer mechanical coupling (Figure 9d).[[3]] This should be further discussed in comparison to the case of DWCNTs, taking into account the difference in thermal expansion coefficients of graphene and h-BN.[[60]] More straightforward signatures can be obtained by Raman spectroscopy. Jeong *et al*. have fabricated highly porous, freestanding heterostructures of SWCNT/h-BN by the pyrolysis of SWCNTs/$B_2O_3$ hydrogel under nitrogen atmosphere at 1000°C. The formation of C-N bonds during the h-BN growth, though undesirable, gives rise to a detectable Raman peak at 1490 cm$^{-1}$ and peaks from off-resonantly excited BN layers are observed even for the aerogel samples (Figure 9e).[142] Since TEM-based identification of the h-BN structures cannot be applied to heterostructures in most devices, it is highly desired to have some optical tools that not only indirectly suggest the existence of h-BN[143] but also provide detailed information, such as the number of layers and their crystallinity. Fluorescence from the color centers of BNNTs[144] would be an affordable option to evaluate their crystallinity of an isolated SWCNT@BNNT heterostructures.

As mentioned above, the outer BN coating could serve as a very homogeneous and non-perturbing protection layer for the CNTs. While excitons in CNTs are known to decay nonradiatively when contacting CNTs with many solid substrates,[145,146] or interacting with molecules in the environment[123], recent studies show that planar h-BN serves as a substrate that does not significantly quench the excitons in SWCNTs.[147,148] This role of h-BN could be extended to the tubular form. Indeed, preliminary studies indicate that bright PL was observed from SWCNTs coaxially wrapped with BNNTs (Figure 9f),[3] which is promising for photonics applications, though a more in-depth study, preferably of heterostructures studied both by TEM and optical spectroscopy, will be required to verify this hypothesis. The chemical stability of h-BN against oxygen at high temperature can be another protective feature that BNNTs can provide.[3,149] Perhaps not surprisingly, CNT samples can be reinforced through CVD-based BN coating on CNT walls and inter-tube junctions, since BNNTs exhibit as excellent mechanical and thermal properties as CNTs. Compressive strength and shape recoverability of CNT forests can be enhanced, owing to the synergistic effect between the CNTs and outer BNNT coating,[149] while thermal conductance of dry-deposited SWCNT films was enhanced without much sacrificing their electrical conductance (Figure 9g).[150]

As graphene-based electronics and physics have been fueled by h-BN as a support, the same should be expected for the 1D heterostructures consisting of CNTs and BNNTs. Several challenges are, however, still to be resolved for device applications of such materials: such as a uniform BN coating over the entire length of the CNTs and a good electrical contact between the metal and the inner CNTs through the BN layers. Moreover, the CVD growth of BNNTs around CNTs requires air-suspended CNTs and substrates that withstand high temperatures. Once uniform BN layers are formed, insulating properties of h-BN (Figure 9h)[3] that depend on its layer number should be controlled by selective etching of BNNTs layers over CNTs[151,152] or other approaches for the use in electronic devices, where a good electrical contact to CNTs is necessary.

In addition to providing inner CNTs with complementary properties due to the presence of the outer h-BN, it may also be possible to add new functionalities through interlayer couplings that could occur under special conditions. Similarly as for DWCNTs, various moiré superlattices can be formed depending on the combinations of chiral angles, tube diameters, and handedness. For example, spiral topological heteronanotubes are theoretically predicted to function as nanoscale solenoids to induce magnetic fields for certain pairs of inner-outer tube chiral indices,[153] though such topological physics could be found in other systems of multi-walled nanotubes. Considering its wide band-gap and atomically thin/flat nature, the more unique and synergetic role that BNNTs can play will be atomic-thick spacers to tune the extent of interlayer coupling between CNTs and other nanocomponents in a ternary structure, as is often the case for 2D materials.[154,155]

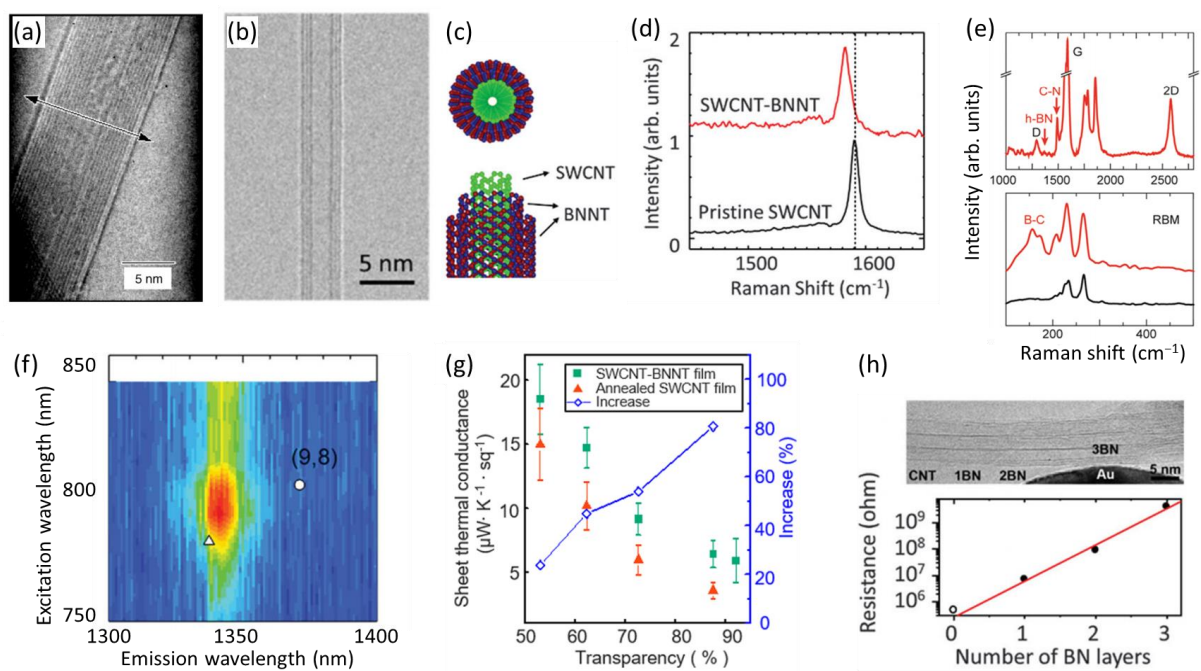

Figure 9. **One dimensional van der Waals heterostructures carbon and hexagonal boron nitride**. (a) TEM image showing the graphitic planes and well-separated BN layers. Panel (a) is produced with permission from [[36]]. Copyright 1997 American Association for the Advancement of Science. (b,c) TEM image (b) and corresponding atomic model (c) of a SWCNT wrapped with two layers of BNNT. (d) Typical G band of an individual SWCNT before and after BN coating. Dotted line indicates the original G band position at ~1590 cm$^{-1}$. (e) Raman spectra from SWCNT@h-BN aerogels showing signatures associated with C–N bonds and h-BN structures (upper panel). RBM spectra taken after the h-BN coating (lower panel). Panel (e) is reproduced with permission from [[142]]. Copyright 2015 Royal Society of Chemistry. (f) PLE map of a suspended (9,8) SWCNT after BN coating. Circle and triangle marks indicate the transition energies of a suspended (9, 8) SWCNT in air and in vacuum,

respectively. (g) Sheet thermal conductance of SWCNT@BNNT films and annealed SWCNT films. The blue hollow rhombuses show the increasing ratio of the thermal conductance after BNNT coating (right axis). Panel (g) is reproduced with permission from [[150]]. Copyright 2020, American Chemical Society. (h) Bright-field TEM image (upper) and electrical resistance versus the number of BN layers (lower). Panels (b-d,f,h) are reproduced with permission (not yet possible) from [[3]]. Copyright 2020 American Association for the Advancement of Science.

## 6. Heterostructures of carbon nanotubes coated by TMDCs

Transition metal dichalcogenides (TMDCs) are layered materials with weak vdW interlayer interactions and strong in-plane bonding, and have unique electronic, optical, chemical, mechanical, and thermal properties.[156] Atomically thin TMDCs have exhibited appealing physics, including 2D superconductivity,[157] valley polarization,[158] and the quantum spin Hall effect,[159] indicating various applications for electronic and optoelectronic devices.[156,160,161] A wide variety of 2D TMDCs have been synthesized from molten-salt-assisted CVD.[162] Similar to the carbon allotropes of graphene and carbon nanotubes, cylindrical structures of TMDCs, known as a family of inorganic nanotubes, also have revealed interesting physical properties for functional devices.[163–165] TMDC nanotubes such as $MoS_2$,[166] $MoSe_2$,[167] $WS_2$,[168] $WSe_2$,[167] and $NbS_2$[169] have been reported. Here, we review recent experimental and theoretical efforts to elucidate optical properties of 1D heterostructures using CNTs as templates for confining and directing the growth of TMDC tubular structures.

## 6.1 Multi-walled carbon nanotubes coated by TMDCs (MWCNT@TMDCs)

Unlike h-BN, a monolayer of TMDC usually has three atomic layers. It is therefore in general more difficult for TMDC materials to form small diameter tubular structures because of their high stiffness. Most efforts on building a 1D $MoS_2$ heterostructure, therefore, started with MWCNTs, whose outer diameters are usually large. Among many TMDCs, molybdenum disulfide ($MoS_2$) has been investigated most intensively in recent decades due to its intriguing physical properties, stability, abundance, and flexibility. For example, it has been shown that the electronic structure of 2D $MoS_2$ would change with the decrease of the number of layers from indirect to direct bandgap.[170] Therefore, a number of studies focused on using MWCNTs as a template to promote the formations of tubular $MoS_2$ around their surface (MWCNT@$MoS_2$). In 2000, MWCNT@$MoS_2$ heterostructures synthesized by pyrolyzing $MoO_3$ coating on MWCNTs in a $H_2S/N_2$ atmosphere at a high temperature were reported.[171] The hydrothermal process, as an important wet chemistry method, allowed to prepare a small number of $MoS_2$ layers on the surface of the MWCNTs. Figure 10a presents a MWCNT@$MoS_2$

heterostructure synthesized by the hydrothermal reaction of $Na_2MoO_4$ and $CS(NH_2)_2$ in the presence of MWCNTs.[172] The tubular $MoS_2$ layers surrounding the surface of the MWCNTs exhibit cylindrical symmetry with decent crystallinity. One potential application of such MWCNT@$MoS_2$ heterostructures is energy storage devices, where the $MoS_2$ nanotube layers provide many active sites and suitable-size channels for lithium-ion intercalation, while the CNTs serve as a conductive network.[173,174] Since the interface electronic states in the heterostructures determine their electronic and optical properties, studies of the electronic structure of MWCNT@$MoS_2$ provide more information about the potential of these heterostructures. Koroteev *et al.* have revealed a downshift (0.15 eV) of the C 1s peak of the MWCNT@$MoS_2$ heterostructure as compared to the pristine MWCNT sample by X-ray photoelectron spectroscopy (Figure 10b) which was attributed to charge transfer between both components.[175] They further confirmed this charge transfer by near-edge X-ray absorption fine structure (NEXAFS) spectroscopy through a reduction of the π*-resonance in the CK-edge NEXAFS spectrum of the heterostructure with respect to the pristine MWCNTs. In addition, the authors speculated that MWCNTs with a defective surface could prevent growth of hexagonal $MoS_2$ layers. Later, optoelectronic applications based on MWCNT@$MoS_2$ heterostructures were proposed, where $MoS_2$ serves as a light sensitizer and the MWCNTs conduct the light-induced charge carriers.[38] Figure 10c shows the photocurrent response of the MWCNT@$MoS_2$ planar device under different laser power intensities. Under laser illumination, electron-hole pairs are generated inside the $MoS_2$ layers after which they are transferred to the inner MWCNTs that then transport the charges, resulting in a positive photoresponse. The device presented a clear linear photoresponse as the power density of the laser changed from 60 to 7.7 mW/cm$^2$, indicating its potential as a photodetector under a wide range of intensities. The increase of the dark current in Figure 10c may be originating from the heating effect on the MWCNTs due to a combined effect of the laser irradiation and the Joule heating. MWCNT@$MoS_2$ heterostructures have thus been proven to be a promising candidate in photodetecting applications.

Besides $MoS_2$, MWCNT templates could also guide the growth of other TMDCs around the MWCNTs. It is noteworthy that niobium disulfide ($NbS_2$) is a layered material possessing intriguing electronic properties (*e.g.,* superconductivity).[176] MWCNT@$NbS_2$ heteronanotubes (as shown in Figure 10d) have been generated via a carbon nanotube template process.[177] To achieve this, the niobium source was initially oxidized and subsequently sulfurized, after which it formed a tubular crystalline structure on the surface of the MWCNTs. The MWCNT@$NbS_2$ heterostructures presented a promising use as a nanocable, with the properties of metallic-metallic or a semiconducting-metallic depending on the electronic state of the inner MWCNTs. By adopting the approach of sheathing MWCNTs with NbS$_2$, rhenium disulfide (ReS$_2$) was successfully coated on MWCNTs (Figure 10e).[178] Unlike other layered TMDCs, ReS$_2$ contains in its ordinary form metal-metal bonded clusters and metal atoms that are octahedrally rather than trigonal prismatically coordinated with sulfur.[179] Figure 10e presents the TEM image of a MWCNT covered with several ReS$_2$ layers. The distance between the ReS$_2$ layers is 0.62 nm which matches the Re-Re interlayer distance in ordinary ReS$_2$ where one Re atom layer and two S atom layers form a sandwiched structure.

Tungsten disulfide (WS$_2$) retains similar electronic properties to MoS$_2$, and its bandgap changes from indirect in bulk (~1.3 eV) to direct (~2.1 eV) in monolayer,[180] enabling applications such transistors and photodetectors. Single-walled TMDC nanotubes such as WS$_2$ nanotubes, if the diameter is not extremely small, will hold similar band structures as their 2D counterparts. Single- and double-walled WS$_2$ layers coated on MWCNTs (as shown in Figures 10f and 10g) were reported by pyrolysing H$_2$S/N$_2$ over WO$_3$-coated CNTs at 900°C.[181] Although the diameter of the produced heterostructure is as large as tens of nanometers, the work implied that MWCNTs could provide a support for topological reactions and the conversion from WO$_3$ to WS$_2$ without creating defects. However, the electronic and optical properties of single-walled WS$_2$ nanotubes were not studied in this work. Following the achievement of producing binary phase WS$_2$-C nanotubes, Walton *et al.* have also attempted to investigate the coating of WS$_2$ nanotubes on SWCNT bundles.[182] Figure 10h shows several SWCNTs encapsulated in WS$_2$ nanotubes, where the inner diameter of WS$_2$ is much larger than the size of the SWCNT bundle. In their work, the conversion from WO$_{3-x}$ to WS$_2$ was proven to occur on the surface of SWCNTs during H$_2$S pyrolysis, which results in the encapsulation of SWCNTs in WS$_2$ nanotubes, although SWCNTs did not seem to serve as the template for the growth of WS$_2$ nanotubes.

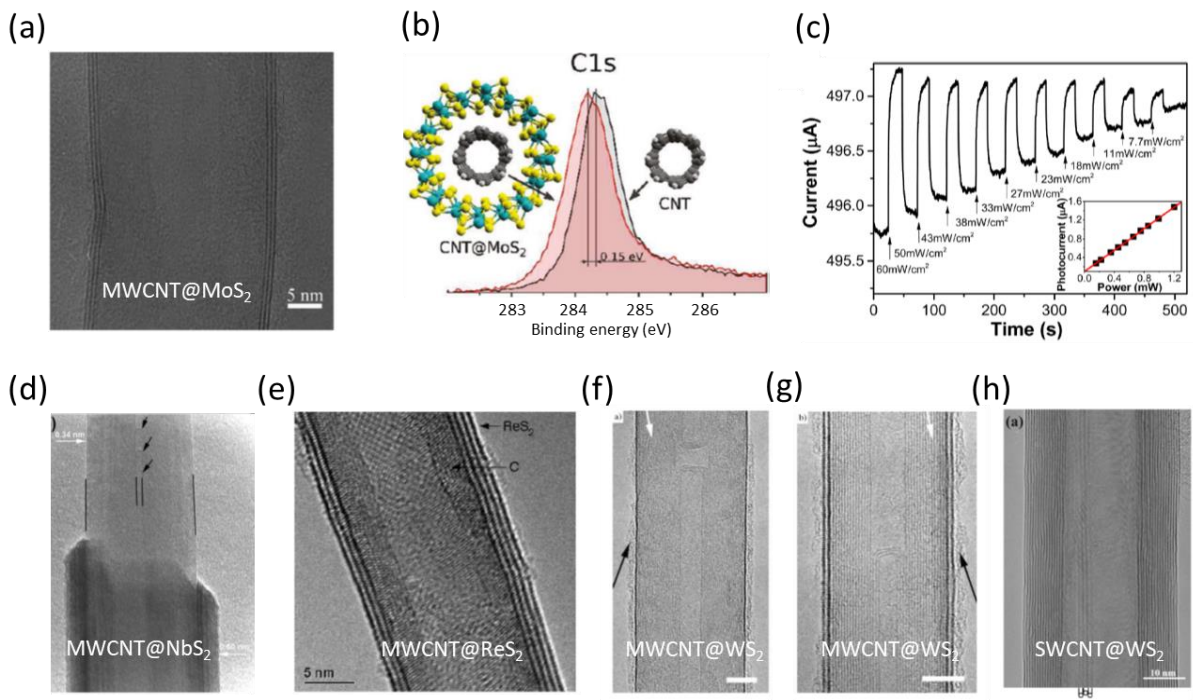

Figure 10. **1D heterostructures of (MW)CNT@TMDCs.** (a) Representative HRTEM image of MWCNT@MoS$_2$ core-shell structure prepared by hydrothermal reaction. Reproduced with permission from [[172]]. Copyright 2006, IOPscience. (b) XPS C 1s spectra of pristine MWCNTs and MWCNT@MoS$_2$ heterostructures, which indicate charge transfer between the MWCNTs and MoS$_2$. Reproduced with permission from [[175]]. Copyright 2011, American Chemical Society. (c) Photocurrent response under different laser power intensities using MWCNT@MoS$_2$ heterostructures. Inset shows the linear dependence of photocurrent on power. Reproduced with permission from [[38]]. Copyright 2015, American Chemical Society. (d) HRTEM image of NbS$_2$ nanotube grown on a close-tipped MWCNT. The outer NbS$_2$ layer distance is 0.6 nm, and the inner carbon layer distance is 0.34 nm. Reproduced with permission from [[177]]. Copyright 2001, Royal Society of Chemistry. (e). HRTEM image of a MWCNT@ReS$_2$ core-shell heteronanotube. Reproduced with permission from [[178]]. Copyright 2002, American Chemical Society. (f,g) HRTEM images of a MWCNT coated with a (f) single and (g) double layer of WS$_2$. Scale bars = 5 nm. Reproduced with permission from [[181]]. Copyright 2001, Wiley-VCH. (h) HRTEM image of a 15 walled WS$_2$ nanotube containing a SWCNT bundle. Scale bar = 10 nm. Reproduced with permission from **[**[182]]. Copyright 2001, American Institute of Physics.

### 6.2 One-dimensional (1D) van der Waals heterostructures of SWCNTs coated by TMDCs

Recently 1D vdW heterostructures based on SWCNTs have been realized by CVD synthesizing boron nitride (BN) and MoS$_2$ shells onto SWCNTs, where the outer layers of BN nanotubes (BNNTs) and MoS$_2$ nanotubes (MoS$_2$NTs) are single-crystalline cylinders.[3] This ternary heteronanotube SWCNT@BNNT@MoS$_2$NT (Figure 11, a-d) has a 5-nm-diameter tubular structure, holding an inner SWCNT, a middle three walls BNNT, and a single outer layer of MoS$_2$NT. The non-catalytic coating method demonstrated by Xiang *et al.* extends the vdW heterostructures from 2D materials to other coaxial materials.[3] Beyond BN and MoS$_2$, also other transition metal chalcogenides and borides, and oxides may be grown by CVD on carbon nanotube cores.[183]

Following this experimental synthesis of single-crystalline 1D vdW heteronanotubes, several experimental studies aimed to investigate the optical properties of 1D heteronanotubes by probing the optoelectronic and photofluorescent signals. Ultrafast optical measurements using pump-probe spectroscopy provide the possibility of observing the charges and excitons in materials. A 1D vdW heterostructure consisting of bundles of SWCNTs sheathed with BNNTs and MoS$_2$NTs (C@BN@MoS$_2$NT) (Figure 11e) has been investigated by ultrafast pump-probe spectroscopy across the visible and terahertz frequency ranges.[184] The transient absorbance in the UV and visible range at different pump-probe delay times after femtosecond pulsed excitation was traced to identify whether excitons in the MoS$_2$NTs are the major photoproduct upon light absorption by MoS$_2$ layers of the heterostructure. Figure 2f presents the fitted transient absorbance in the range near the A and B excitons from C@BN@MoS$_2$NT heteronanotubes. Excitons are the primary photoproduct in the MoS$_2$ nanotubes of the C@BN@MoS$_2$NT heteronanotube because A and B excitonic absorption was present at all time delays. The broadening of the A exciton after 1 ps results from faster energy, phase, or momentum relaxation after photoexcitation, similarly as in atomically thin 2D TMDCs.[185] The increase of the A-exciton strength after 100 ps is unusual, which may originate from the dynamic screening of the Coulomb interaction or from trion formation. Optical pump-THz probe (OPTP) spectroscopy could examine the photoconductivity and detect the presence of free charges in semiconductor nanomaterials.[186],[187] The negative differential transmission $\Delta T/T$ corresponding to a positive photoconductivity at later times in Figure 11g reveals the mobile charges in the MoS$_2$NTs in C@BN@MoS$_2$NTs. The negative photoconductivity at early times was attributed to optical excitation of the SWCNTs and the dynamic switch from negative to positive reflects the differing temporal dynamics for free-carrier absorption in SWCNTs and MoS$_2$NTs.

Later, the PL from single-walled MoS$_2$NTs coaxially grown on BNNTs was observed experimentally (shown in Figure 11h).[188] Similar to monolayer 2D MoS$_2$,[170] single-walled MoS$_2$NT shows a strong PL emission while the PL from multi-walled MoS$_2$NT would be significantly quenched. The PL of single-walled MoS$_2$NT from SWCNT@BNNT@MoS$_2$NT heteronanotubes was also found to be quenched in Figure 11h, which might result from the charge transfer and energy transfer between SWCNT and MoS$_2$NT (through BNNT) in the heteronanotube. The DFT calculation performed in this work presented a bandgap crossover from an indirect to a direct band structure when the chiral index of the single-walled MoS$_2$NT approaches (30,30), for which the diameter of the nanotube is about 5.2 nm (as shown in Figure 11i). The DFT result is consistent with the observed PL signal from single-walled MoS$_2$NT in the experiment. Together they evidenced that single-walled MoS$_2$NTs with large diameter are direct bandgap semiconductors.

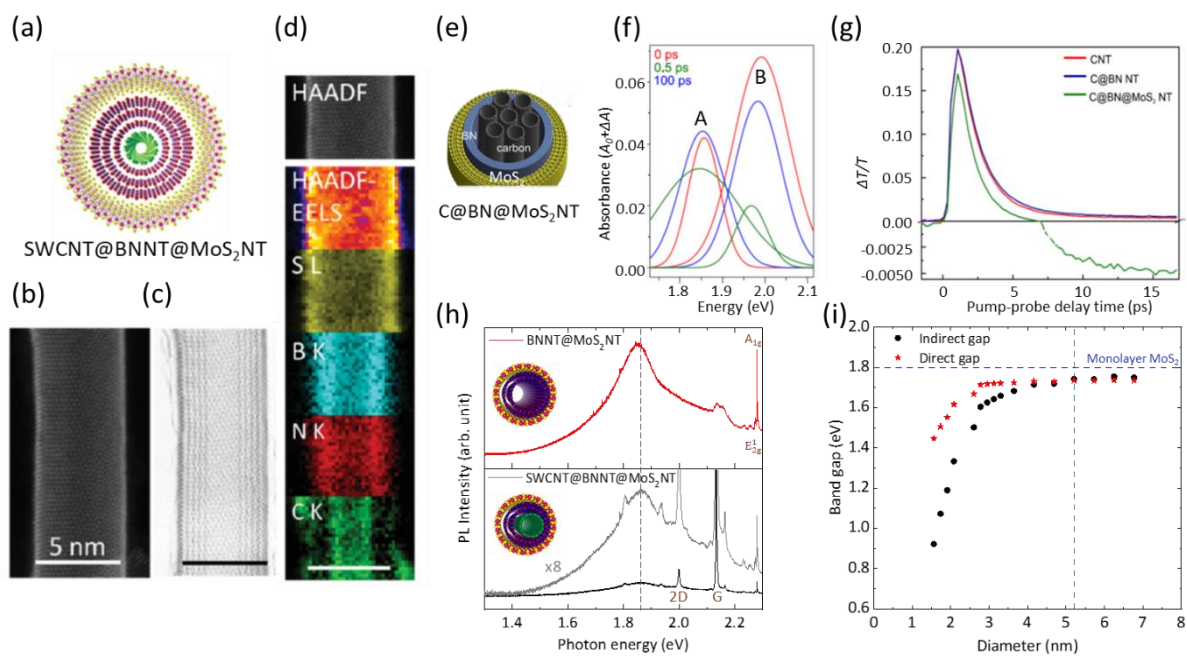

Figure 11. **Single-crystalline 1D SWCNT based heterostructures.** (a) Atomic model, (b) HAADF-STEM image, (c) annual bright field (ABF)-STEM image, and (d) EELS mapping of a 5-nm-diameter ternary SWCNT-BNNT-MoS$_2$ 1D vdW heterostructure. Scale bars = 5 nm. (a-d) Reproduced with permission from [[3]]. Copyright 2020, American Association for the Advancement of Science. (e) Atomic model of C@BN@MoS$_2$NT heterostructure (bundled SWCNT as a template). (f) Fitted transient absorbance in the range near the A and B excitons of MoS$_2$. Transient absorption spectra as a function of pump−probe delay time. (g) Transient THz response of CNT, C@BNNT, and C@BNNT@MoS$_2$. (e-g) Reproduced with permission from [[184]]. Copyright 2020, American Chemical Society. (h) The PL spectra of BNNT@MoS$_2$NT and SWCNT@BNNT@MoS$_2$NT heteronanotubes with atomic modes inset. (i) The bandgap values of armchair MoS$_2$NT as a function of diameter. (h-i) Reproduced with permission from [[188]]. Copyright 2021, American Chemical Society.

## 6.3 Heterostructures based on CNTs coated by oxides and beyond

Beyond a single type of TMDC, tubular structures of transition metal oxides as well as multiple types of TMDC heterostructures can be fabricated. One example is the coating of CNTs with single-crystalline and single-walled molybdenum trioxide ($\alpha$-MoO$_3$) nanotubes (MONTs) (as shown in Figure 12, a-c).[189] The chirality of the MONTs can be directly recognized from high-angle annular dark field (HAADF)-STEM images. The epitaxial growth with curvature-induced deformations and dislocations presents more complex geometric selectivity in terms of intrinsic lattice matching. Moreover, inorganic core-shell nanotubes that combine two different TMDC materials $WS_2@MoS_2$ have been achieved by direct chemical reaction on $WS_2$ nanotubes (Figure 12d).[190] Because of two layers of TMDC in a 1D heterostructure are curved in different radius, their coupling will be different, e.g. moiré patterns, from the case of 2D. If smaller diameter nanotubes can be synthesized, 1D quantum confinement is also expected to occur. However, optical and electronic properties of the core-shell $WS_2@MoS_2$ nanotubes were not investigated in this work. In addition, coaxial 1D vdW heterostructures comprising a CNT core and a thickness-tunable covalent organic framework (COF) shell (CNT@COF) were very recently reported. This CNT@COF 1D heterostructure was applied as a bifunctional oxygen electro-catalyst in rechargeable zinc–air batteries, delivering a high specific capacity and excellent cycling stability.[191]

The attempt of assembling core-shell single-walled TMDC nanotubes might be a promising research direction for novel optoelectronics and light-harvesting devices. Atomically 2D thin $MoS_2/WS_2$ heterostructures have been theoretically and experimentally confirmed to form a type II heterojunction (Figure 12e),[192] with the conduction band minimum (CBM) residing in $MoS_2$ and the valence band maximum (VBM) in $WS_2$. Upon excitation, electron-hole pairs are initially created in the $MoS_2$ layer, but holes transfer quickly to the $WS_2$ layer while electrons stay in the $MoS_2$ layer, thereby forming interlayer excitons in the heterostructure. The charge transfer between the two layers in the 2D heterostructure was evidenced by transient absorption spectroscopy, as presented in Figure 12f. Similar to 2D monolayer TMDC heterostructures, single-walled 1D TMDC heteronanotubes could potentially be expected to form a type II band alignment or a type III broken-gap band alignment.[193] Besides, a theoretical work has predicted that SWCNTs and DWCNTs could possess flexoelectricity, because the curvature of the nanotube breaks the mirror symmetry of the $p_z$-orbitals compared to graphene (Figure 12g-h).[194] In a DWCNT, a significant charge transfer from the outer tube to the inner tube due to the hybridization of the orbitals located on different walls was presented in this work (as shown in Figure 12i). Based on these as-mentioned theoretical predictions,[192–194] cylindrical heterostructures that combine different semiconducting single-walled TMDC materials or semiconducting SWCNTs with single-walled TMDCs preserve an intriguing potential for electronic and optoelectronic devices. In addition to the 1D radial junction, which corresponds to a vertical junction in a 2D heterostructure, axial junctions of two different TMDC nanotubes, or even axial superlattices, may also be possible to fabricate considering the successful growth of in-plane 2D heterostructures.[195] This demonstrates that there is plenty of room for nanotube-based 1D heterostructures to stimulate many research interests from material synthesis to device applications.[196]

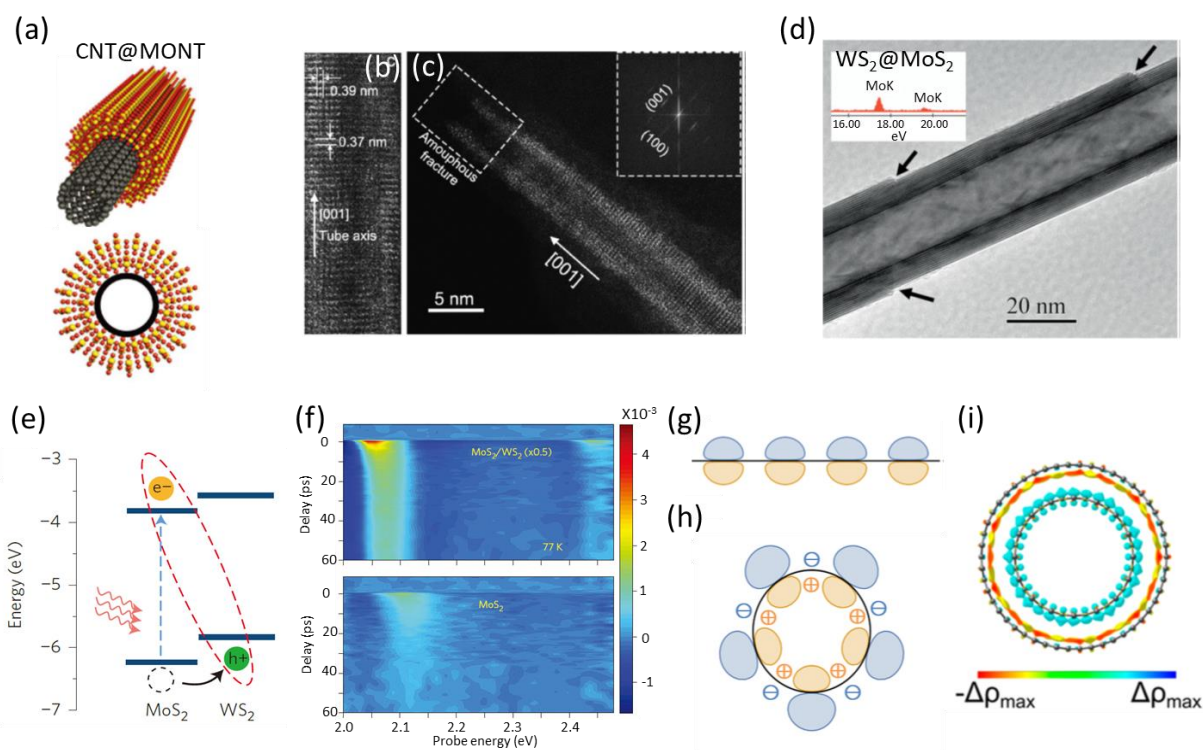

Figure 12. **Single-crystalline CNT@MONT, TMDC@TMDC, and beyond.** (a) Atomic structure of CNT@MONT. (b) HAADF-STEM image of MONT. (c) HAADF-STEM image of the fracture of MONT. The uniform fringes and pattern in the inset FFT image provides the lattice information of MONT. (a-c) Reproduced with permission from [[189]]. Copyright 2018, Wiley-VCH. (d) TEM image of a $WS_2$@$MoS_2$ core-shell NTs prepared via a gas-phase reaction. The inset is a part of the EDS spectrum, exhibiting Mo peaks. Arrows represent $MoS_2$ layer. Reproduced with permission from [[190]]. Copyright 2010, Wiley-VCH (e) Schematic of the theoretically predicted band alignment of a 2D $MoS_2$/$WS_2$ heterostructure, a type II band alignment. (f) Transient absorption spectra from a 2D $MoS_2$/$WS_2$ heterostructure and an isolated $MoS_2$ monolayer upon excitation of $MoS_2$ A-exciton transitions. (e-f) Reproduced with permission from [[192]]. Copyright 2020, Springer Nature. The symmetry of $\pi$-orbitals in graphene (g) is broken in nanotubes by their curvature, displacing electron density outward and creating a potential difference across the wall (h). (i) Isocharge plot showing significant charge transfer from the outer tube to the inner tube. (g-i) Reproduced with permission from [[194]]. Copyright 2020, American Chemical Society.

## 7. Summary and outlook

This review summarized the recent progress on building various nanoscale heterostructures using (mainly single-walled) carbon nanotubes as a template. A particular emphasis was placed on the various fabrication approaches as well as the resulting optical properties of these 1D heterostructures. Entirely, the activities so far can be divided into inner approaches and outer approaches, with the former utilizing the SWCNT inner space as a container and the latter using the SWCNT outer surface as a substrate. While the heterostructures synthesized by encapsulation of molecules inside the SWCNTs' hollow core have been studied in detail in the past years, and many examples can be found in literature, the large variety in the choice of the molecule combined with the various SWCNT chiralities as containers, still provides a gigantic, unexplored playing ground to create and tune new functionalities. In contrast, the efforts for making heterostructures by crystallized layers on the outer surface of the SWCNT are still at an infancy stage in terms of both the number of materials that have been studied/tried and the variety of properties that have been characterized.[150,197,198]

DWCNTs, which can be considered as a homo-atomic, yet hetero-structural 1D textbook example, have been intensively studied in the past decade. Their vdW-coupling-induced properties are best understood among all different 1D heterostructures. The interlayer electron coupling, mechanic-deformation-related phonon-coupling and moiré-induced new states are well demonstrated by both theoretical and experimental studies. Filling different molecules, like dyes or solvents, into the channel of a SWCNT provides large freedom to tune the intrinsic SWCNT properties, *e.g.*, by shifting their optical transitions, together with an opportunity to study the properties of these molecules in a confined nanospace, such as phase transitions and specific diameter-dependent 1D stacking. With the aid of heat or electron beam irradiation, encapsulated molecules and atoms can even transform into novel structures that cannot be easily prepared in free space. Ultra-thin carbon or boron-nitride nanotubes, linear chains of carbon or other atoms, 1D nanowires and ultranarrow graphene nanoribbons are representative examples. Although most literature studies on coating the outer surface of CNTs to obtain heterostructures have only resulted in SWCNT@amorphous or MWCNT@crystal heterostructures, highly crystalized BNNT and $MoS_2$NT on SWCNT, *i.e.*, SWCNT@crystal 1D vdW heterostructures have recently been achieved. In this case, the outer $MoS_2$ is curved with a diameter of only 4-5 nm, close to the minimum predicted by theory. When the diameter is large, the properties of the outer $MoS_2$ should be close to that of the corresponding 2D layers, even though it is in a tubular shape, but in such small-diameter heterostructures, the curvature effect and the confinement in the radial direction are expected to have their impact, giving rise to interesting phenomena that could be more complicated than those being intensively discussed in 2D vdW heterostructures.

Despite the quick and growing progress, there are still many questions that are yet to be answered and many interesting research directions that are yet to be explored. We outline some of potential topics that we believe will appear in the near future.

1. Although DWCNTs are the most-studied 1D heterostructures, a large number of varieties in the inner@outer tube combinations have complicated the precise understanding of the tube-tube interaction. If one considers handedness in addition to the interlayer distance and chirality of the DWCNTs, the variety of structural combinations increases even further. Clarifying the handedness could lead to better understanding and tuning of the properties of DWCNTs, such as interlayer moiré excitons. One possible strategy for a next-stage experiment could employ a chirality specific SWCNT,[199,200] sorted by DGU or other approaches,[201,202] as a starting material, and synthesize an inner SWCNT by thermal conversion of precursor molecules, as presented in [[103]], strongly reducing the number of possible inner tubes that can be synthesized by limiting the outer tube to a single chirality. This might provide a great convenience for understanding DWCNTs, the simplest and textbook 1D heterostructure that one can imagine so far.

2. The intrinsic properties of many types of novel structures formed inside the nanochannel of SWCNTs are not fully explored yet. Part of the reason is that most of these experiments are done using SWCNTs as the container, which, *e.g.,* completely quench the emission of the encapsulated species, due to the typically much smaller bandgap for the host SWCNTs as compared to the encapsulated structures. One solution to this is using the larger band gap BNNTs as an alternative, as recently demonstrated in [[99]]. Its 6 eV band-gap will allow a non-perturbing environment for most optical characterizations. However, to achieve similar 1D heterostructures as currently prepared for SWCNTs, holding just a single or double file of dye molecules, small-diameter, long and isolated BNNTs will be highly desired for these studies.

3. Another challenge to solve is to extract inner structures, *i.e.,* nanotubes, nanowires, graphene nanoribbons, linear carbon chains, from the channel into the free space without damaging these inner structures. While inner nanotubes are typically extracted by sonication,[103,106,203] the sonication is known to cut the tubes into short pieces and create a lot of defects. When applying this technique to linear carbon chains (LCC) encapsulated inside DWCNTs (LCC@DWCNTs), intending to obtain LCC@SWCNT samples, the Raman signal of the LCCs was greatly reduced, indicative of the destruction of most of the encapsulated LCCs[204], in particular when taking into account its giant Raman cross-section.[118] A different strategy that does not destroy the newly synthesized structures is thus highly required. Even without the sonication, material stability is another concern and still unknown. Indeed the ultralong LCCs might not be stable in the free space, but at least graphene nanoribbons and atomic WTe nanowires have been synthesized in the bulk phase and have been proven to be stable.[205]

4. The study of 1D vdW heterostructures built on the outer surface of SWCNT is still in its early stage. The crystal synthesis methodology on the SWCNT's tiny and highly curved surface is not fully established yet, and the number of materials demonstrated is still very limited. Moreover, their electronic properties are barely explored. [41][141] In principle, the outer surface of a SWCNT can provide a larger space/freedom for the construction of a variety of heterostructures. Different radial heterostructures, which correspond to vertical structures in the 2D case, could be designed, enabling the formation of more types of semiconductor-semiconductor junctions.[143] An axial junction that corresponds to a planar heterojunction in 2D materials should also be possible. This can possibly give a ultrasmall diode with a diameter of only several nanometers. Of course, the experimental challenges in synthesis and characterizations of such tiny tubular structures can be very high, but benefiting from the development of SWCNTs, nowadays ultralong SWCNTs are available that can be suspended over a trench or deposited on a substrate.[206] These ultralong SWCNTs may serve as a perfect starting material for synthesis, characterization, and property investigation for 1D vdW heterostructures. The fundamental crystal growth behaviors, *e.g*., nucleation mechanism and growth dynamics, on this curved surface with a radius of only around 1 nm, need to be further explored and understood.

**Acknowledgements**

Part of this work was supported by JSPS KAKENHI (grant numbers JP18H05329, JP19H02543, JP20H00220, and JP20KK0114) and by JST, CREST grant number JPMJCR20B5, Japan. Part of this work was supported by the Fund for Scientific Research Flanders (FWO) through postdoctoral grant of D.L. (12ZP720N) and the European Research Council (Starting Grant no. 679841, ORDERin1D). This work is also supported by a JSPS-FWO bilateral joint research project (grant number FWO: VS08521N).